%% file: main.tex
\pdfoutput=1

\documentclass[journal,cosmoc]{IEEEtran}
\usepackage{amsmath,amsfonts}
\usepackage{array}
\usepackage{subcaption}
\usepackage{textcomp}
\usepackage{stfloats}
\usepackage{url}
\usepackage{verbatim}
\usepackage{graphicx}
\usepackage{cite}
\usepackage[outline]{contour}
\contourlength{1pt}

\input{macros.tex}

\input{tikz_style.tex}

\input{symbols.tex}

\input{glossaries}

\begin{document}

\title{Word Error Rate Definitions and Algorithms for Long-Form Multi-talker Speech Recognition}

\author{Thilo~von~Neumann, Christoph~Boeddeker, Marc~Delcroix, Reinhold~Haeb-Umbach
\thanks{Thilo von Neumann and Christoph Boeddeker contributed equally.}
\thanks{Christoph Boeddeker was funded by Deutsche Forschungsgemeinschaft (DFG), project no. 448568305.}
\thanks{Thilo von Neumann, Christoph Boeddeker, and Reinhold Haeb-Umbach are
with the Paderborn University, 33098 Paderborn, Germany (e-mail: vonneumann@nt.upb.de; boeddeker@nt.upb.de; haeb@nt.upb.de).}
\thanks{Marc Delcroix is with NTT Inc., Communication Science Laboratories, Kyoto 619-0237, Japan (e-mail:
marc.delcroix@ieee.org).}
}

\maketitle

\begin{abstract}

The predominant metric for evaluating speech recognizers, the Word Error Rate (WER) has been extended in different ways to handle transcripts produced by long-form multi-talker speech recognizers.
These systems process long transcripts containing multiple speakers and complex speaking patterns so that the classical WER cannot be applied.
There are speaker-attributed approaches that count speaker confusion errors, such as the concatenated minimum-permutation WER cpWER and the time-constrained cpWER (tcpWER), and speaker-agnostic approaches, which aim to ignore speaker confusion errors, such as the Optimal Reference Combination WER (ORC-WER) and the MIMO-WER.
These WERs evaluate different aspects and error types (e.g., temporal misalignment). 
A detailed comparison has not been made.
We therefore present a unified description of the existing WERs and highlight when to use which metric.
To further analyze how many errors are caused by speaker confusion, we propose the Diarization-invariant cpWER (DI-cpWER).
It ignores speaker attribution errors and its difference to cpWER reflects the impact of speaker confusions on the WER.
Since error types cannot reliably be classified automatically, we discuss ways to visualize sequence alignments between the reference and hypothesis transcripts to facilitate the spotting of errors by a human judge.
Since some WER definitions have high computational complexity, we introduce a greedy algorithm to approximate the ORC-WER and DI-cpWER with high precision ($<0.1\%$ deviation in our experiments) and polynomial complexity instead of exponential.
To improve the plausibility of the metrics, we also incorporate the time constraint from the tcpWER into ORC-WER and MIMO-WER, also significantly reducing the computational complexity.
\end{abstract}


\glsresetall

\section{Introduction}

The predominant evaluation metric for speech recognition systems is the \gls{WER} \cite{huang2001a,Watanabe2020_CHiME6ChallengeTackling,povey2011kaldi,watanabe2018espnet,ravanelli2021speechbrain}.
It was proposed for single-speaker single-utterance recognition where the system output, the hypothesis, is compared to a single ground-truth reference transcript \cite{Young1998_SpeechRecognitionEvaluation,povey2011kaldi}.
In its basic form, it counts the minimum number of substitutions, deletions, and insertions of a recognizer's output relative to the reference transcript, divided by the number of words in the reference.

With the current shift in research towards long-form multi-talker speech recognition \cite{Watanabe2020_CHiME6ChallengeTackling,Chen2020_ContinuousSpeechSeparation,vinnikov2024notsofar,yu2022summary}, this simple definition reaches its limits \cite{sklyar2022multiturn,vonNeumann2022_WordErrorRate}.
With long-form multi-talker speech recognition we refer to answering the question who spoke when (diarization) and what was said (speech recognition) for a recording of arbitrary length.
Within one recording, multiple speakers interact with frequent speaker turns and pauses.

Today's systems answer these questions to varying degrees, sometimes neglecting diarization, speaker attribution, temporal information, or all of them.
There are systems that perform both diarization and transcription and provide speaker labels and temporal annotations with the transcript \cite{Cornell2023_CHiME7DASRChallenge,raj2021integration,Boeddeker2023_TSSEPJointDiarization}.

These systems often comprise sub-systems that target smaller sub-problems and produce various output formats.
\gls{CSS} systems, for example, have a fixed number of output streams, but the transcripts of different speakers can be mixed in the same stream \cite{Chen2020_ContinuousSpeechSeparation,yoshioka2018recognizing}.
\gls{SOT} \cite{Kanda2020_SerializedOutputTraining,kanda2022streaming} systems may even output only a single stream where special speaker change tokens are inserted into the transcript to speaker changes.
These systems are to some degree free to place the recognized speech on any output stream so that the standard \gls{WER}, which requires a single reference transcript to compare with, cannot be computed.
There can be ambiguities, e.g., in speaker labels and hypothesis-to-reference assignments.

These ambiguities must be solved before a WER can be computed, which is usually done by assigning reference utterances to system outputs or vice versa \cite{Watanabe2020_CHiME6ChallengeTackling,sklyar2022multiturn,Raj2022_ContinuousStreamingMultiTalker,vonNeumann2022_WordErrorRate}.
Different approaches exist for finding the desired assignment; no single solution is applicable in all cases.

One example of a widely known speaker-attributed \gls{WER} for long-form transcripts is the \gls{cpWER} \cite{Watanabe2020_CHiME6ChallengeTackling}.
It assumes that the system groups its output by speakers and the recognized transcripts, i.e., the concatenation of several utterances, of each speaker are compared to the reference transcripts of each speaker.
Only the \gls{WER} for the best-matching mapping is reported.
When evaluation is desired with a time constraint to improve the plausibility of the matching, the \gls{tcpWER} \cite{vonNeumann2023_MeetEvalToolkitComputation} emerges from the \gls{cpWER} when words that are temporally far apart in the reference and hypothesis are never matched as correct or substitution.

The \gls{ORC-WER} \cite{sklyar2022multiturn,Raj2022_ContinuousStreamingMultiTalker,vonNeumann2022_WordErrorRate} is a speaker-agnostic \gls{WER} definition.
Instead of solving a permutation problem across speakers, each reference utterance is assigned to one output stream, such that all assignments together minimize the error rate.
A more permissive generalization of \gls{ORC-WER} is the \gls{MIMO-WER} \cite{vonNeumann2022_WordErrorRate} that allows the global temporal order to be violated while maintaining the order of utterances from the same speaker.

What is missing among this variety of \gls{WER} definitions in the literature is a systematic comparison that highlights their differences.
Currently, different WERs have been proposed in various publications, with different definitions, making their comparison difficult. 
We introduce a unified formalization, which clearly shows their differences and similarities.
With this formalization, we introduce a taxonomy of WER approaches and compare them theoretically and experimentally. 
We present examples that illustrate the differences and facilitate choosing an appropriate definition for a use case.

To judge the impact of speaker attribution errors on the \gls{cpWER}, we propose the \gls{DI-cpWER}, which corrects speaker assignment errors such that the error rate is minimized.
The DI-cpWER is the value of the cpWER if the estimated speaker labels were correct.
The difference between DI-cpWER and cpWER measures the number of errors caused by speaker confusions.

Additionally, we propose a greedy algorithm to solve the assignment problem for the DI-cpWER and ORC-WER, which often yields the optimal result while providing a significant speedup in many cases.
This further broadens the applicability of the metrics to more complex scenarios.

While the WERs can give a good picture of the overall performance of a recognizer, they lack providing detailed insights into the types of errors a system has made.
Heuristics can be found for classifying errors in more detail, but we refrain from using them because different error types cannot always be discerned reliably.
Instead, we argue that a visualization of the alignments computed by the WERs is more valuable in practice.
We describe a way of interactively visualizing these sequence alignments such that a user can easily spot errors.
A tool to generate these interactive visualizations is available in the MeetEval toolkit (see \cref{sec:viz,sec:open-source}).

In the remainder of the paper, we first describe the types of errors that can occur in long-form multi-talker speech recognition in \cref{sec:error-types}.
After introducing our notation in \cref{sec:notation}, we overview the \gls{WER} definitions found in the literature, using our unified notation.
In \cref{sec:time-constrained}, we describe the time-constrained word error rates.
In \cref{sec:diarization-invariant}, we discuss the impact of diarization on the \gls{WER} and introduce the \gls{DI-cpWER} to estimate the impact of speaker attribution errors on the \gls{cpWER}.
We further propose a greedy approach for computing the ORC-WER and DI-cpWER.
As a tool for developing systems, we present visualization methods for sequence alignments in \cref{sec:viz}.
We conclude with experiments comparing \glspl{WER} and algorithms.

\subsection{Contributions}

This work builds upon our prior works which propose the MIMO-WER, an efficient exact algorithm to compute the MIMO-WER and ORC-WER \cite{vonNeumann2022_WordErrorRate} and the time-constrained minimum-permutation WER (tcpWER) \cite{vonNeumann2023_MeetEvalToolkitComputation}.

In this work:
\begin{itemize}
    \item We introduce a unified formalization of the different WERs found in the literature and discuss their differences and similarities with the aim to provide guidance as to which metric to use when.
    \item We describe in detail the Diarization-invariant cpWER (DI-cpWER) for system analysis as a means to estimate the impact of speaker attribution errors on the cpWER.
    \item We integrate the time-constraint from tcpWER \cite{vonNeumann2023_MeetEvalToolkitComputation} into the MIMO-WER \cite{vonNeumann2022_WordErrorRate} and ORC-WER \cite{sklyar2022multiturn,Raj2022_ContinuousStreamingMultiTalker,vonNeumann2022_WordErrorRate}, leading to more realistic alignments and a significant computational speedup.
    \item We propose a greedy algorithm for (tc)ORC-WER and (tc)DI-cpWER that closely approximates the exact solution while significantly reducing computation.
    \item We present a visualization tool for the analysis of alignments in long-form transcripts and show its usability in a case study.
\end{itemize}

\subsection{Related work}

\subsubsection{Alternative metrics}
    
Other metrics have been proposed for measuring transcript quality that define different ways of translating the matching obtained by the Levenshtein algorithm into an error measure, such as the Match Error Rate (MER) \cite{Morris2004_WERRILMER}, Relative Information Lost (RIL) \cite{Miller1955NoteOT} or Word Information Lost (WIL) \cite{Morris2002_InformationTheoreticMeasure}.
The MER aims to create a symmetric measure that is bound by \SI{100}{\percent} for easier interpretability.
However, symmetry is not always desired, and being bound by \SI{100}{\percent} limits the number of errors the metric can measure.
The RIL and WIR aim to measure the information lost by the errors made by the transcription system instead of counting the raw word errors.
These metrics, however, make assumptions about relations between words (RIL measures zero errors for any one-to-one mapping between reference and hypothesis words) and are not as simple to compute as the WER.
Nevertheless, the RIL, WIR, and MER, similar to the WER, are based on the Levenshtein alignment of reference words to hypothesis words, so the approaches discussed in the remainder of this work are also applicable to them.

Yet other approaches evaluate the transcript content rather than the words.
Approaches like semantic distance (SD) \cite{Kim2021_SemanticDistanceNew}, BERTScore \cite{Zhang2019_BERTScoreEvaluatingText} or $\text{H}_\text{EVAL}$ \cite{sasindran2023heval} extract embeddings that represent semantic meaning either on a word or transcript level with a pre-trained model and compute the distance between them.
$\text{H}_\text{EVAL}$ combines SD with a keyword-error-rate.
These metrics depend on a pre-trained (neural network) model, which makes their computation and interpretation difficult.
Numbers are only comparable when the same model is used for all systems, which is not always feasible.

The \gls{DER} is also frequently used for meeting-level systems.
It only evaluates speech activity and speaker attribution estimates by measuring the ratio of the duration of wrongly recognized speech activity to the ground truth speech activity.
Its collar option defines temporal regions around the start and end points of reference utterances that are not scored in order to account for the fact that the beginning and end points of speech are not clearly defined.
The collar option in the time-constrained WER variants (\cref{sec:time-constrained}) is inspired by the \gls{DER}'s collar.
Its effect is, however, different. 
While the \gls{DER} ignores errors within the range of the collar around the start and end points of a segment, (so $\text{DER}=0$ for a collar $c \rightarrow \infty$), the time-constrained WERs exclude any words from matching as correct or substituted when they are farther apart than the collar (i.e., they approach their non-time-constrained variants, e.g., $\text{tcpWER}\overset{c\rightarrow\infty}{=}\text{cpWER}$).
Also, unlike in the DER, a non-zero collar is often required since word-level annotations are often unavailable.
Instead of judging full segments (as DER), the WERs look at a word level, where small temporal deviations have a much larger impact than on a segment level.

All algorithms described in this work can also be applied on a character level, resulting in the \gls{CER}, or be used to compute a \gls{SER}.

\subsubsection{Related tools}

The NIST Scoring Toolkit (SCTK)\footnote{\url{https://github.com/usnistgov/SCTK/}} contains tools for single- and multiple-sequence alignment.
The single-sequence alignment provided by \verb|sclite| contains rudimentary time constraints.
It splits the word sequences where no word is active and scores the regions in between with the standard (non-time-constrained) WER.
The multi-sequence alignment provided by \verb|asclite| employs similar concepts to the word-level (time-constrained) MIMO-WER (\cref{sec:mimo}).
It uses a multi-dimensional alignment algorithm, presented in \cite{Fiscus2006_MultipleDimensionLevenshtein}, and requires word-level time stamps.
In contrast to the MIMO-WER, the asclite tool does not have the utterance consistency constraint, making it over-optimistic.

The Kaldi toolkit \cite{povey2011kaldi} provides tools for single-speaker WER computation and basic plain-text-based alignment visualization.
A variety of other implementations exist for single-speaker WER, e.g., in SpeechBrain \cite{ravanelli2021speechbrain} or jiwer\footnote{\url{https://github.com/jitsi/jiwer}}.
These implementations do either not apply to long-form speech recognition or are specialized and highly integrated into a framework.

\section{Errors in Long-Form Meeting Recognition and human interpretation}
\label{sec:error-types}

The performance of long-form multi-talker speech recognition systems is typically evaluated with the \gls{WER} \cite{huang2001a,Watanabe2020_CHiME6ChallengeTackling,watanabe2018espnet,ravanelli2021speechbrain,Morris2004_WERRILMER,Young1998_SpeechRecognitionEvaluation}, either speaker-attributed or speaker-agnostic, and the \gls{DER} \cite{NIST2009_2009RT09Rich}.
The WER measures word-level errors (word insertions, deletions, and substitutions), while the DER assesses diarization errors, i.e., temporal misalignment and speaker attribution errors.

\subsection{Error types}

In long-form meeting recognition, errors can occur that are typically unobserved in simpler scenarios.
We here discriminate between recognition, activity, speaker attribution and segmentation errors.

Recognition errors are errors that are solely related to the speech content.
These errors are the only errors measured by the WER in single-utterance recognition scenarios.
They include word insertions, word deletions and word substitutions.

Activity errors are errors where the predicted speech does not match the ground-truth speech activity.
These errors include missed speech, where speech is present in the reference but no transcript is predicted and false alarm, where a transcript is predicted where no speech is present in the reference.
A special case of false alarm is leakage, where a transcript of a reference utterance is predicted on multiple output streams, often with degraded performance.

When more than one speaker is active and speaker attributions are predicted, errors can also appear in the predicted speaker attribution.
Speaker attribution errors include any wrongly assigned speaker labels, e.g., a single wrong speaker label or swapped labels (speaker confusions).

Segmentation errors appear when the segmentation, i.e., the grouping of words into utterances, differs between reference and hypothesis.
An utterance split divides a single reference utterance into multiple hypothesis segments while an utterance merge groups multiple reference utterances into a single hypothesis segment.
Both can appear simultaneously.
Segmentation errors alone are typically not measured by the WER, but only when they are paired with speaker attribution errors.
This is desired since the ground-truth segmentation is often arbitrary.

None of these errors are measured directly by \gls{WER} metrics.
They are only reflected in the measured number of word insertions, word deletions and word substitutions.
Any activity and speaker recognition errors would ideally result only in word insertions and deletions (missed speech: a sequence of deletions; wrong speaker label: insertions for one speaker and deletions for another speaker).
But, because the WER minimizes the number of errors, insertions and deletions are usually merged into substitutions where possible.
They are additionally mixed up with recognition errors.
It is thus impossible to reliably identify these higher-level errors from the word-level errors alone.

Some \gls{WER} definitions try to eliminate substitution errors caused by high-level errors by either ignoring them (see speaker-agnostic metrics in \cref{sec:speaker-agnostic-wers}) or by splitting them into insertions and deletions (by a time-constraint described in \cref{sec:time-constrained}).
We still refrain from classifying the higher-level error types automatically.
As an alternative, we provide a visualization tool in which a human can identify these types of errors and analyze the system (\cref{sec:viz}).

\subsection{Human interpretation of errors}
\label{sec:human-interpretation}

Although humans tend to have a good intuition for judging transcription errors, it highly depends on the visualization of the transcripts and may be different for different humans.
Translating human intuition into a metric is thus difficult.

We focus only on the intuitively recognized errors that can be reliably algorithmically detected. 
Namely, we introduce a temporal constraint in \cref{sec:time-constrained} to model intuition on the physical properties of speech more closely.
We also present the DI-cpWER in \cref{sec:diarization-invariant} as a means of indirectly quantifying speaker attribution errors, which are often intuitively easy to detect by a human but impossible to reliably classify algorithmically from the transcript alone.

We do not consider human perception of these errors, i.e. their impact on the perceived quality of the transcript.
Some errors certainly impact the transcript quality more than others; missing filler words have low impact while a missing negation may alter the meaning of the transcribed text.
Instead, we focus on the definition and implementation of WERs for meetings that we believe are intuitively reasonable for several application scenarios. 
A more rigorous analysis of the correlation between perceived transcript quality and the different types of WER is out of the scope of this paper.

\section{Notation}
\label{sec:notation}

We assume that all transcripts are provided as segments, where a segment is a continuous region of speech activity of a single speaker.
The reference and hypothesis segments are represented by vectors of segment indices $\rv=[1,2,3,...]\tran\in\mathbb{N}^{\dim(\rv)}$ and $\hv=[1,2,3,...]\tran\in\mathbb{N}^{\dim(\hv)}$, where $\dim(\rv)$ and $\dim(\hv)$ are the numbers of segments in the reference and hypothesis, respectively.
Every reference and hypothesis segment has a transcript and a (reference speaker or system output stream) label.
The transcripts are stored in vectors $\rtrn = [\rtrne_1, ...]\tran$, $\htrn = [\htrne_1, ...]\tran$, 
where each element is a word sequence.
The speaker or stream labels are also vectors $\rlbl = [\rlble_1,...]\tran\in\{1,...,\nspk\}^{\dim(\rv)}$ and $\hlbl=[\hlble_1,...]\tran\in\{1,...,\nstream\}^{\dim(\hv)}$.
Vectors marked with a superscript $()^\text{ref}$ or $()^\text{hyp}$ have the same dimension as $\rv$ and $\hv$, respectively.

If the system predicts no speaker labels (e.g., a \gls{CSS}-style system \cite{yoshioka2018recognizing,Chen2020_ContinuousSpeechSeparation}), the system's output stream index is used as a label.
The number of reference labels (speakers) is denoted by $\nspk$ and the number of hypothesis labels (system output streams or estimated speaker labels) is denoted by $\nstream$.
If the system produces timestamps, the begin and end times of the segments are $\tbeg=[\tbege_1,...]$ and $\tend=[\tende_1,...]$.
We assume that the segments in $\rv$ and $\hv$ are initially sorted by ascending begin time $\tbeg$ and in this work we call this the global temporal ordering.

We define a \emph{selection vector} as a multi-hot vector $\vect{v}\in\{0,1\}^{\dim(\rv)}$ that selects a subset of segments from $\rv$.
We use the notation $\ind_{\{i:\mathrm{condition}(i)\}}$ for the selection vector that selects all segments $i$ for which $\mathrm{condition}(i)$ is true.
The \emph{selection matrix} $\select(\vect v)\in\{0,1\}^{N\times\dim(\rv)}$ is a matrix constructed from a selection vector $\vect v$ by deleting those rows from the identity matrix of size $\dim(\rv)$ that correspond to the zeros in $\vect{v}$, e.g.,
\begin{align}
    \select([1,0,0,1]\tran) = \left[\begin{matrix}
        1&0&0&0 \\
        0&0&0&1
    \end{matrix}\right].
\end{align}
Multiplying such a matrix to a vector results in a vector of dimension $N=||\vect{v}||^2$ (the number of ones in $\vect{v}$) that contains only the selected entries of the original vector in the original order: $\select([1,0,0,1]\tran)\cdot[1, 2, 3, 4]\tran = [1, 4]\tran$.
Note that a selection matrix has a \enquote{step structure}: all entries right and above a $1$ or left and below a $1$ are $0$.

We define a \emph{selection reorder matrix} as a matrix $\selection \in \{0,1\}^{N\times\dim(\rv)}$ that allows selection and reordering.
It has the same shape as a selection matrix but does not necessarily have the step structure.
It can be constructed by multiplying a selection matrix with a permutation matrix.

We denote with $\vect1_D$ and $\vect0_D$ vectors of dimension $D$ filled with ones or zeros, respectively.

\section{A Survey of Word Error Rate Definitions}
\label{sec:survey}

This section describes the standard WER and existing extensions of the WER for long-form meeting recognition.
We provide a (new) unified mathematical description of different approaches with which we aim to align each metric with human intuition and the issues a human would find.

\subsection{Purposes for using a metric}

Before discussing the WERs themselves, it is important to understand what metrics are typically used for.

A metric is often used to rank systems across (independent) works.
In this use case, the metric should correlate with the (total) system performance.
It specifically should be \emph{unfoolable}, i.e., there should be no obvious way to improve the metric's value without improving the actual system performance.

Another application is system analysis, where the goal is to analyze specific aspects of a system's output.
For this purpose, metrics often have to be specific, for example the number of speaker confusion errors, and may not fulfill the unfoolable condition.
This is because the researcher who analyzes a system is typically interested in improving the system performance and not (only) the metric.

    Further complexity is added because different applications may favor different errors.
    For live captioning, for example, speaker errors are less important because a listener can often see or hear which speaker is speaking.
    For a summary task it is important that the meaning of the transcript is preserved, but filler words are mostly unimportant.
    This means that different metrics suit different scenarios.

Given the complexity of the errors and human interpretation (\cref{sec:error-types}) with the different ways to apply the metrics, we can already conclude that there is no single ideal metric for long-form meeting recognition.
Instead, one or multiple metrics have to be selected individually for every use-case.
Our descriptions in the following aim to convey the properties of each metric to facilitate the selection process.

\subsection{Levenshtein distance}
\label{sec:lev}

All WER metrics are based on the Levenshtein distance.
We assume for its definition that every segment constitutes a single word  for simplicity without loss of generality.
See \cref{sec:sec:pseudo-word-level-timestamps} for transitioning from segment-level annotations to word-level annotations.
The Levenshtein distance $\lev(\rv,\hv)$ is the (total) minimum number of word insertions $I$, word deletions $D$ and word substitutions $S$, required to transform one word sequence $\rv$ into another word sequence $\hv$.
It can be found with the Wagner-Fisher algorithm \cite{Wagner1974_StringtoStringCorrectionProblem}, which recursively computes the Levenshtein distance as
\begin{align}
    \lev(\rv,\hv, C_\text{S},C_\text{I},C_\text{D}) =
        \min \begin{cases}
             \lev(\rv,\select(\vect{v}_{\hv})\hv) + C_\text{D},\\
             \lev(\select(\vect{v}_{\rv})\rv,\hv) + C_\text{I},\\
             \lev(\select(\vect{v}_{\rv})\rv,\select(\vect{v}_{\hv})\hv)\\ \phantom{++} + C_\text{C/S}(r_{\rv},h_{\hv},C_\text{S}),
        \end{cases}
        \hspace{-.25em}
        \label{eq:lev}
\end{align}
where the selection vector $\vect{v}_{\rv}=[1,...,1,0]\tran{\in\mathbb{R}^{\dim(\rv)}}$ contains $\dim(\rv)-1$ ones and one zero to select all elements but the last.
The algorithm is initialized with $\lev(\rv,[]\tran)=\dim(\rv)$ and $\lev([]\tran,\hv)=\dim(\hv)$.
The costs for insertion and deletion are set to $C_\text{I}=C_\text{D} = 1$.
The cost for a match $C_\text{C/S}$ depend on the words, where a correct match (same word) does not increase the distance ($C_\text{C}=0$) while a substitution (different word) increases the distance by $C_\text{S}=1$:\footnote{Other choices are possible when computing the matching. Sometimes $C_\text{I}=C_\text{D}=4$, $C_\text{S}=3$ and $C_\text{C}=0$ is used to prefer substitutions over insertions and deletions. Any weights with $C_\text{C}<C_\text{S}\leq C_\text{I} = C_\text{D}$ are allowed\cite{Morris2004_WERRILMER}, but may produce different matchings.}
\begin{align}
    C_\text{C/S}(r, h, C_\text{S}) = \begin{cases}
        0, & \text{if } \rtrne_r = \htrne_h,\\
        C_\text{S}, & \text{otherwise.}
    \end{cases}
    \label{eq:lev-cost}
\end{align}
We neglect the costs in the arguments of $\lev$ and $C_\text{C/S}$ for brevity where possible.
The alignment can be reconstructed as a sequence of edit operations by backtracking through the matrix that stores the intermediate distances.

Note that this matching is plausible for short recordings, but may not be plausible when matching long sequences containing natural speech (as discussed later in \cref{sec:time-constrained}).

\subsection{Standard utterance-wise WER}
With the notation from \cref{sec:notation} and $\cref{eq:lev}$, the utterance-wise \gls{WER} the ratio of the Levenshtein distance, $\lev$, and the total number of words, $N^\text{ref}$, in the reference:
\begin{align}
    \mathrm{WER} 
    = \frac{\sum_{n}\lev(\rv_{n},\hv_{n})}{{\sum_{n}}N^\text{ref}_{n}} 
    = \frac{{\sum_{n}}I_{n}+D_{n}+S_{n}}{{\sum_{n}}N^\text{ref}_{n}}.
    \label{eq:wer}
\end{align}
The sum goes over all examples from a dataset where $\rv_n$ and $\hv_n$ come from the same example.
The example index $n$ is neglected in the following for brevity.
As shown in the latter part of the equation, $\lev(\rv,\hv)$ can be decomposed into the number of insertions $I$, deletions $D$, and substitutions $S$ required to transform $\rv$ into $\hv$.
Note that this decomposition is not unique, i.e., multiple different decompositions can lead to the same (minimal) WER.

\paragraph{Strengths} The standard \gls{WER} definition is well known, generally accepted and widely used. It has a relatively simple interpretation as the ratio of wrongly transcribed words to the total number of words.

\paragraph{Weaknesses} By being such a simple definition, the standard \gls{WER} can only be applied to the simplest of all scenarios: a single utterance. 
Anything more complex, be it multiple speakers or long sequences where temporal alignment plays a larger role, needs a more sophisticated WER definition.

\paragraph{Application} 
Since the standard WER can only be computed between exactly two word sequences, it is well suited for evaluating conventional single-speaker single-utterance recognizers.
It can be applied to a system that transcribes multiple speakers when the true speaker identity is known (i.e., there is no ambiguity).
If the true speaker identity is unknown, it is common to choose the best assignment (compare \cref{sec:cpWER}).

\subsection{Meeting-level Word Error Rates}

\begin{figure*}
    \input{wer-definition-figure}
    \caption{Example assignments for different WERs using the \emph{trace} visualization (see \cref{sec:viz}). Letters represent words, boxes represent segments. WERs are computed as $\text{WER}=\frac{I+D+S}{N^\text{ref}}$.}
    \label{fig:wer-visualization}
\end{figure*}

Recently, the development of speech recognition systems has shifted towards more complex scenarios with long recordings, multiple speakers, multiple utterances, and more complex output formats.
In these scenarios, the standard WER can no longer be computed, due to ambiguities in the matching between the reference and the hypothesis transcripts, e.g., a permutation across speaker labels.
Different approaches from the literature for solving the ambiguity are discussed in the remainder of this section.

All WERs described below can be traced back to \cref{eq:wer} except that the ambiguity of assigning reference segments to system output streams (or vice versa) is solved beforehand.
We here only describe the modified distance functions that replace $\lev$ in \cref{eq:wer} for the different definitions.
\cref{fig:wer-visualization} lists examples of solving the assignment problem for the different WERs, where the assigned reference or hypothesis after solving the ambiguity is labeled as \emph{modified} reference or hypothesis.
Pseudo-code-like implementations of all algorithms in Python are available at \url{https://github.com/fgnt/meeteval/blob/main/doc/algorithms.md} as supplementary material.

\subsection{Speaker-attributed WERs}
Speaker-attributed \glspl{WER} take speaker attribution (i.e., estimated speaker labels) into account.
The simplest and most widely used one is the \gls{cpWER}.

\subsubsection{cpWER}\label{sec:cpWER}
\glsreset{cpWER}
The \gls{cpWER} \cite{Watanabe2020_CHiME6ChallengeTackling} assumes that the system groups its output by speakers, but the true speaker identity is unknown.
It concatenates all transcripts of the same speakers and computes the minimum WER across all possible permutations between reference and hypothesis speakers, as indicated in \cref{fig:example-cp}.
It thus minimizes the distance w.r.t. a one-to-one mapping (or, if $\nspk=\nstream$, a permutation) $\perm^\text{cp}$ between reference and hypothesis labels $\rlbl$ and $\hlbl$:
\begin{align}
    \distance^\text{cp} &= \min_{\perm} \sum_{\ispk=1}^\nspk \lev(\select(\ind_{\{i:\rlble_i = \ispk\}})\rv, \select(\ind_{\{i:\hlble_i=\pi_\ispk\}})\hv).
    \label{eq:cp}
\end{align}
The expression $\select(\ind_{\{i:\rlble_i = \ispk\}})$ creates a selection matrix (see \cref{sec:notation}) that selects all segments from speaker $\ispk$.
$\nspk$ is the number of reference speakers.
If the number of reference and hypothesis labels are different, dummy labels with empty transcripts are inserted until the reference and hypothesis have the same number of labels so that every reference transcript is matched with exactly one hypothesis transcript.

\paragraph{Strengths}
The \gls{cpWER} is widely known and accepted for scenarios with a limited number of speakers but long recordings, such as the data provided in the CHiME challenges \cite{Watanabe2020_CHiME6ChallengeTackling,Cornell2023_CHiME7DASRChallenge}. 
It penalizes not only word errors, but also speaker assignment errors.
It is unaffected by utterance splits and merges, as long as the speaker label is not modified.

\paragraph{Weaknesses} 
When used for systems that perform diarization (i.e., segmentation \emph{and} speaker attribution) along transcription, it only judges speaker attribution errors and no temporal errors.
If temporal alignment is important and such an evaluation is desired, the \gls{tcpWER} can be used (see \cref{sec:time-constrained}).

\paragraph{Application}
The \gls{cpWER} is well suited for systems that estimate speaker identities along with the transcript, but not necessarily perform temporal segmentation.

\subsubsection{DA-WER}\label{sec:dawer}
In an attempt to find a metric that measures both diarization and recognition performance for the 7th CHiME challenge, the \gls{DA-WER}\cite{Cornell2023_CHiME7DASRChallenge} was defined.
It is similar to the \gls{cpWER}, but the permutation $\perm^\text{DA}$ minimizes the \gls{DER} instead of the \gls{WER}:
\begin{align}
    \distance^\text{DA} &=  \sum_{\ispk=1}^\nspk\lev(\select(\ind_{\{i:\rlble_i = \ispk\}})\rv, \select(\ind_{\{i:\hlble_i=\pi^\text{DA}_\ispk\}})\hv),\\
    \perm^\text{DA} &= \argmin_{\perm}  \sum_{\ispk=1}^\nspk \text{DER}(\select(\ind_{\{i:\rlble_i = \ispk\}})\rv, \select(\ind_{\{i:\hlble_i=\pi_\ispk\}})\hv).
\end{align}
Note that the permutation does not depend on the transcription, but only on the diarization result.

\paragraph{Strengths} This metric forces systems to produce a somewhat meaningful diarization along the transcripts.

\paragraph{Weaknesses}
While the permutations $\perm^\text{DA}$ and $\perm^\text{cp}$ found with \gls{DER} and \gls{WER} are frequently equal, this is not guaranteed.
The permutation found by the \gls{DER} may be sub-optimal for \gls{WER} if speakers have a similar temporal activity (extreme case: full overlap).
The assignment only has a minor impact on the \gls{DER} if the speakers have similar activity but it may arbitrarily impact the \gls{WER}.

\paragraph{Application} 
The \gls{DA-WER} can be applied to all systems that perform diarization and transcription, but long recordings are necessary for reliable results.

\subsection{Speaker-agnostic WERs}
\label{sec:speaker-agnostic-wers}

The goal of speaker-agnostic \glspl{WER} is to ignore speaker attribution errors.
Since these cannot easily be fully discriminated from other error types (see \cref{sec:error-types}), they are usually not fully invariant to speaker label errors.

\subsubsection{ORC-WER}
\label{sec:orc}
The \gls{ORC-WER} \cite{sklyar2022multiturn,Raj2022_ContinuousStreamingMultiTalker} is designed for \gls{CSS}-style systems where each stream contains a set of segments that may come from different speakers, i.e., the system does not attribute segments to speakers.
Therefore, the most plausible assignment of reference utterance onto output streams has to be found.
This is done by modifying the reference labels, $\rlbl$, such that the  \gls{WER} is minimized under the following \emph{utterance-consistency-constraint}:
An utterance must be mapped completely on a (single) system output stream and may not be split across streams.
Additionally, the global utterance order is kept.
\cref{fig:example-orc} shows the result of this optimization for one example.
Compared to the cpWER, every utterance receives a label instead of every speaker.

The optimization is here written as minimizing the distance over selection vectors $\vect v_1, ..., \vect v_\nstream$ (see \cref{sec:notation}) that select the reference utterances for each stream:
\begin{align}
    \distance^\text{ORC} = \min_{(\vect v_1, ...)\in \mathcal V} \sum_{\istream=1}^\nstream \lev(\select(\vect v_\istream)\rv, \select(\ind_{\{i:\hlble_i=\istream\}})\hv).
    \label{eq:orc}
\end{align}
The minimum is computed over the set
{$\mathcal{V}=\{(\vect v_1, ..., \vect v_\nstream) : \vect v_\istream \in \{0,1\}^{\dim(\rv)}, \sum_{\istream=1}^\nstream \vect v_c = \vect1_{\dim(\rv)}\}$} of all
$|\mathcal V| = \nstream^{\dim(\rv)}$ tuples of $\nstream$ selection vectors that sum to $\vect1_{\dim(\rv)}$.
The constraint on $(\vect{v}_1, ..., \vect{v}_\nstream)$ ensures that every segment is selected exactly once.
The selection matrices (see \cref{sec:notation}) are constructed such that the global temporal order of segments is not changed, but only the assignment of reference segments to system output streams.

\paragraph{Strengths}
The ORC-WER penalizes utterance splits (i.e., when a system splits an utterance of a single speaker across multiple outputs), but no (other) speaker attribution errors.
It can, in that sense, be seen as diarization-invariant.

\paragraph{Weaknesses}
The main drawback of the ORC-WER is its high computational complexity.
In rare cases, the ORC-WER can become higher than expected.
This happens when the system changes the order of utterances, which may occur when they are temporally close or for certain ASR network architectures such as SOT \cite{Kanda2020_SerializedOutputTraining,kanda2022streaming}\footnote{The MIMO-WER is designed for SOT to address the weakness of the ORC-WER for SOT systems.}.
Such an edge-case is displayed in \cref{fig:example-orc}, where the \enquote{D} and \enquote{F} are swapped by the system and \gls{ORC-WER} is unable to match the (correctly transcribed) \enquote{D} between reference and hypothesis.

\paragraph{Application}
The \gls{ORC-WER} can be applied to systems produce multiple output streams but do not perform speaker attribution, e.g., the \gls{CSS} pipeline \cite{Chen2020_ContinuousSpeechSeparation}, where the system outputs a fixed number of overlap-free streams, but the streams do not carry information about speaker identity.
It can also be used to partially evaluate a pipeline, e.g., before the speaker assignment stage.

\subsubsection{MIMO-WER}
\label{sec:mimo}

The MIMO-WER  \cite{vonNeumann2022_WordErrorRate} is a generalization of \gls{ORC-WER} that also allows the global temporal order to be violated:
In ORC-WER a unique global temporal order is used.
In MIMO-WER the ordering is weakened to a partial ordering, where the constrained on the order between speakers is removed, i.e., only the temporal order within a speaker must be kept by the system.
This is visible in \cref{fig:example-mimo}, where the utterances \enquote{D} and \enquote{EF} are reordered on the reference side to achieve a lower MIMO-WER than ORC-WER in \cref{fig:example-orc}.
This is here written as a minimization over selection reorder matrices $\selection$ (see \cref{sec:notation}):
\begin{align}
    \distance^\text{MIMO} &= \min_{(\selection_1, ..., \selection_\nstream)\in\mathcal S} \sum_{\istream=1}^\nstream \lev(\selection_\istream \mathbf r, \select(\ind_{\{i:\hlble_i=\istream\}})\mathbf\hv) ,\label{eq:mimo}
\end{align}
where the set $\mathcal S$ contains all tuples of $\nstream$ selection reorder matrices that fulfill the following constraints.
Every segment is selected exactly once across all $\selection_\istream$, so $[\selection_1\tran,\dots,\selection_\nstream\tran]\tran$ is a permutation matrix.
Additionally, the order of the segments after the assignment must be compatible with the partial (temporal) ordering of segments within each speaker.
This means that there must exists a global ordering of all segments that is consistent with each speaker's ordering, as selected by $\select(\ind_{\{i:\rlble_i=k\}})\rv$, and each assigned stream's ordering, as selected by $\selection_\istream\rv$.

This condition ensures a plausible order of the assigned segments: assignments that cannot be represented with the same underlying global ordering are excluded.
For example, given $\rv = [1, 2, 3, 4]\tran$, $\rlbl = [1, 1, 2, 2]\tran$ and $\rtrn=[\texttt{a}, \texttt{b}, \texttt{x}, \texttt{y}]$, speaker $1$ has utterance \texttt{a} followed by \texttt{b} and speaker $2$ has utterance \texttt{x} followed by \texttt{y}.
The assignment $\selection_1\rv=[2, 3]\tran$ (transcription \texttt{b x}) and $\selection_2\rv=[4, 1]\tran$ (transcription \texttt{y a}) is implausible.
Here, {\texttt{a}} comes before {\texttt{b}} (in the reference), {\texttt{b}} comes before {\texttt{x}} (in the assignment), {\texttt{x}} comes before {\texttt{y}} (in the reference), so by transitivity {\texttt{a}} comes before {\texttt{y}}, which contradicts the second assignment where {\texttt{y}} comes before {\texttt{a}}.

The \gls{ORC-WER} discussed in the previous section is a special case of the MIMO-WER where the utterances of all speakers are merged to form a single reference, i.e., $\rlbl=\vect0_{\dim(\rv)}$ before applying \cref{eq:mimo}.
The \gls{MIMO-WER} is in most cases equal to the \gls{ORC-WER}, except for rare cases when the system changes the order of utterances or when the utterance order is not uniquely defined.

\paragraph{Strengths}
The MIMO-WER is relatively permissive regarding diarization errors and only penalizes utterance splits and order changes within a speaker.
This is explicitly required when evaluating some kinds of system, such as \gls{SOT} systems that may change the temporal order of utterances.

\paragraph{Weaknesses}
Similarly to the ORC-WER, the MIMO-WER has a high computational complexity which prevents it from being used in some scenarios with more than a few speakers, e.g., CHiME or Libri-CSS.
It can also be too permissive since there are fewer constraints on the temporal placement of transcripts.
Both issues can be solved by adding a temporal constraint on the distance for correct matches (see \cref{sec:time-constrained}).
The constraint on temporal order could be undermined by giving each utterance a unique speaker label.
This lowers the final \gls{WER} but dramatically increases the computational complexity so that it cannot be exploited in practice.

\paragraph{Application}
The MIMO-WER can be applied in even more scenarios than the ORC-WER.
It is well suited when a system does not keep the global temporal order of utterances and does not assign speaker identities or an evaluation of the speaker attribution is not desired.
One example of such a system is a t-SOT system that sorts transcripts by speakers before sorting temporally.
The computational complexity is reduced when the system only outputs a single stream.

\section{Time-constrained Error Rates}
\label{sec:time-constrained}
\glsreset{tcpWER}

As already mentioned in \cref{sec:human-interpretation}, a human judge respects the temporal alignment between the words uttered in a recording and the corresponding transcript, while the WER, as described above, does not.
A time-constraint was already mentioned in \cite{Young1998_SpeechRecognitionEvaluation} for the \gls{WER}, but it was disregarded because it provided no benefit for their studies on short recordings.
In \cite{vonNeumann2023_MeetEvalToolkitComputation}, the \gls{tcpWER} was introduced, which prevents unreasonable matchings with a time-constraint when working with \emph{long} word sequences.
The following sub-sections give an overview of why a time-constraint improves the WER in meeting scenarios and discuss the what metrics result from applying the time-constraint to the \glspl{WER} defined above.

For long recordings, the alignment found by the standard Levenshtein algorithm, i.e., the one that minimizes the error rate, can become implausible.
It always prefers a substitution over pair of insertion and deletion, independent of their temporal position.
Such a match is preferred even if the two words are temporally so far apart that a correct matching is practically implausible.
We argue that in these cases, \emph{two} errors should be counted (insertion and deletion) instead of \emph{one} (substitution) or \emph{none} (correct match).
A human annotator would certainly count these matchings as an error and any system that reasonably reflects the physical behavior would not be penalized by such a time-constraint.

\subsection{Time-constrained Levenshtein distance}
\label{sec:tclev}

The Levenshtein distance $\lev$ (\cref{eq:lev}) used in the WER definitions above can be replaced with the time-constrained Levenshtein distance $\tclev$ that forbids correct matches or substitutions across long temporal distances.
The time-constrained Levenshtein distance $\tclev$ is computed similarly to $\lev$ (\cref{eq:lev}) with the only modification that $C_\text{C}=C_\text{S}=\infty$ if words are farther apart than a predefined collar $\collar$ \cite{vonNeumann2023_MeetEvalToolkitComputation}.
The transition costs no longer only depend on the words $\rtrn$ and $\htrn$, but also on their begin and end times $\rtbeg$, $\rtend$, $\htbeg$ and $\htend$ (compare \cref{eq:lev-cost}):
\begin{align}
    C_\text{C/S}^\text{tc}(r, h) = \begin{cases}
        \infty, & \text{if } \rtbege_r - \htende_h >\collar \vee \htbege_h - \rtende_r > \collar, \\
        0, & \text{if } \rtrne_r = \htrne_h,\\
        1, & \text{otherwise.}
    \end{cases}
    \hspace{-.1em}
    \label{eq:tc-cost}
\end{align}

The (exact) computation of $\tclev$ can be sped up by pruning the search space where matches are disallowed by the time constraint.
It is enough to compute a bounded band of the Levenshtein matrix where $C_\text{C/S}^\text{tc}(r, h) \neq \infty$.
The computational complexity thus becomes linear in the number of words in the reference and hypothesis, assuming that the number of words that overlap with each other is limited.

\subsection{Time-constrained word error rates}

Substituting $\tclev$ (\cref{eq:tc-cost}) in the \glspl{WER} from \cref{sec:survey} leads to the following metrics, where we only describe the difference to their non-time-constrained counterpart:

\subsubsection{tcpWER}
Using $\tclev$, instead of the Levenshtein distance $\lev$ in the \gls{cpWER} in \cref{eq:cp} leads to the \gls{tcpWER}.

\paragraph{Strengths} 
The time constraint eliminates the issues from cpWER that matchings across arbitrarily long distances, and with it arbitrary segmentation errors, can still lead to a perfect score.
Compared to the \gls{DA-WER}, the \gls{tcpWER} looks at the transcription and diarization jointly so that timestamps attached to the words must match the actual word positions.

\paragraph{Weaknesses} 
The \gls{tcpWER} introduces the collar as a hyperparameter $\collar$ that must be set, and be kept constant across evaluations to achieve comparable results.
Additionally, the timestamps in the reference must be accurate enough (which can be compensated for with the collar).

\paragraph{Application} 
The \gls{tcpWER} can be applied to any system that provides diarization (segmentation and speaker attribution) along with transcription.

\subsubsection{time-constrained MIMO-WER and time-constrained ORC-WER}
Substituting \cref{eq:tc-cost} in the \gls{ORC-WER} and \gls{MIMO-WER} in \cref{eq:orc} and \cref{eq:mimo} leads to the \gls{tcORC-WER} and \gls{tcMIMO-WER}, respectively.

\paragraph{Strengths} 
Matchings across arbitrary temporal distances are eliminated, similar to the tcpWER. 
The time-constrained versions of the MIMO and ORC-WER allow for a large speedup in computation (see \cref{sec:benchmark}).

\subsection{Pseudo-word-level timestamps}
\label{sec:sec:pseudo-word-level-timestamps}

Until now, we assumed segment-level timestamps in \cref{sec:survey} and word-level timestamps in \cref{sec:tclev}.
A standard way of obtaining word-level timestamps is a forced alignment obtained from a speech recognizer.
The effort to obtain an alignment depends the recognizer architecture and can include (costly) explicit modeling or an addition decoding pass \cite{Hu2025_WordLevelTimestamp}.
When word-level time stamps are not available from the transcription system or the reference annotations, \emph{Pseudo} word-level timestamps can be estimated from segment-level timestamps.
\cite{vonNeumann2023_MeetEvalToolkitComputation} provides several methods for this, the most realistic one being to divide the segment-level annotation into word-level sub-segments where their length is proportional to the number of characters in a word.
This roughly models word pronunciation length \cite{vonNeumann2023_MeetEvalToolkitComputation,cornell2024model}.

To prevent fooling the metric, the hypothesis should use the center-\emph{points} of these character-based timestamps.
Otherwise,  a system can improve the metric by splitting off single words from a segment, e.g., the first and last word, and extending their duration to reduce the probability of deletions.

\section{Diarization-Invariance for Word Error Rates}
\label{sec:diarization-invariant}

For a deep analysis of the errors made by a recognition system it is important to discriminate different error types which is in general impossible, as discussed in \cref{sec:error-types}.
As a proxy for the impact of diarization errors, we may wish to know what would the WER be if the system diarized correctly?
None of the presented WERs is able to do this precisely.

All WERs presented penalize speaker attribution errors and temporal errors to some degree.
Even the speaker-agnostic tcORC-WER and tcMIMO-WER penalize utterance split errors when different sections of a single utterance get assigned different speaker labels.
But for none of the presented error rates, it is clear how many errors are exactly caused by speaker attribution or segmentation errors.

\subsection{Speaker attribution-agnostic WER}

We aim to design a novel speaker attribution-agnostic WER that is compatible to the speaker-attributed cpWER such that the number of errors introduced by speaker confusions can be extracted indirectly by the difference of the two.
For this, we propose to \enquote{fix} the speaker assignment, i.e., modify the hypothesis speaker labels $\hlbl$, such that the WER is minimized, and to compare the resulting WER with the cpWER.

We call this WER computation scheme with and without a time constraint the \gls{DI-tcpWER}  and \gls{DI-cpWER}, respectively.
The name is taken from \cite{Boeddeker2023_TSSEPJointDiarization}, which used an algorithm similar to the greedy algorithm described in \cref{sec:greedy}.
Here, we introduce an exact algorithm to compute the DI-tcpWER and DI-cpWER.

The \emph{Diarization-Invariant cpWER} and \emph{diarization-invariant tcpWER} can be computed by swapping reference and hypothesis in the \gls{ORC-WER} (\cref{eq:orc}) or \gls{tcORC-WER}, respectively while keeping the denominator.
The assignment is solved on the hypothesis side (i.e., estimated labels are modified, or \enquote{corrected}) instead of the reference, as depicted in \cref{fig:example-di-cp}:
\begin{align}
    \distance^\text{DI-cp} = \min_{\vect v_1, ...} \sum_{\istream=1}^\nstream \lev(\select(\ind_{\{i:\rlble_i=\istream\}})\rv, \select(\vect v_\istream)\hv).
    \label{eq:di-cp}
\end{align}
The optimization is the same as ORC-WER in \cref{eq:orc} but with the roles of reference and hypothesis swapped.
Note that no general relation can be stated between ORC-WER and DI-cpWER; utterance merges affect the DI-cpWER negatively and have no effect on ORC-WER, while utterance splits affect DI-cpWER positively and ORC-WER negatively.

\paragraph{Strengths} 
The difference between \gls{DI-cpWER} and \gls{cpWER} can be interpreted as the error that is caused by wrong speaker assignments.
Note that this is not an exact computation, but a rough estimate, because a speaker confusion can have arbitrary effects on the Levenshtein computation.

\paragraph{Weaknesses} 
This way of achieving invariance to speaker attribution errors has significant drawbacks and potential for malicious exploitation, so these metrics should never be used for system ranking.
The metric favors smaller segments.
Its value can be improved by splitting segments into words and letting the metric assign speaker labels on a word level to minimize the \gls{WER}.
This is, however, not a problem for analysis purposes, where the metric is only used to estimate the errors for a single system.
The \gls{ORC-WER} does not suffer from this issue.

\paragraph{Application} 
The \gls{DI-cpWER} is useful for analyzing a system that would otherwise be evaluated with the \gls{cpWER}. 
It is the lower bound on the \gls{cpWER}.
Compared to the ORC-WER, the DI-cpWER counts utterance-merge errors but ignores utterance-split errors.

\subsection{Word-level Diarization-invariant WER}
For a (truly) diarization-invariant WER, errors in the segmentation and speaker attribution tasks must both be ignored.
This can be achieved by dropping the utterance-consistency constraint and solving the assignment problem on a word level instead of the utterance level, similar to what has been done in the NIST toolkit's \verb|asclite| tool \cite{Fiscus2006_MultipleDimensionLevenshtein}.
This means that every word is considered a segment on its own.
Applied to the ORC-WER, MIMO-WER and DI-cpWER, this leads to the word-level ORC-WER (wl-ORC-WER), the word-level MIMO-WER (wl-MIMO-WER), and the word-level DI-cpWER (wl-DI-cpWER).
See \cref{fig:example-wl-orc} for an example of the wl-ORC-WER and \cref{fig:example-wl-di-cp} for wl-DI-cpWER.
Note that this bears disadvantages: the metric is almost guaranteed to under-estimate the number of errors and find unreasonable mappings.
The computation is infeasible for large transcripts.
We here refrain from using and analyzing these word-level metrics, but we mention them in \cref{fig:wer-visualization} to give an idea of how they compare to the other metrics.

\subsection{Estimating speaker confusion errors from speaker-agnostic metrics}

An alternative for estimating the number of speaker confusion errors is to compute the assignment using a speaker-agnostic metric (e.g., tcORC-WER or DI-tcpWER) and count the number of words for which the speaker label changed.
This is similar to the approach used in the (ambiguously) so-called speaker-attributed WER (SA-WER) from NIST \cite{NIST2009_2009RT09Rich}, but can be applied to any speaker-agnostic metric.

This approach is also not exact and incompatible with the cpWER, i.e., the cpWER cannot be computed from the tcORC-WER and the number of speaker confusions.
When computed from a word-level metric, the number of speaker confusions is typically over-estimated.

\section{A Greedy Approximation to DI-cpWER and ORC-WER}
\label{sec:greedy}

We propose to use a greedy approximation when DI-cpWER or ORC-WER are not feasible to compute in a reasonable time.
Here, we describe the algorithm for the DI-cpWER, but it is also applicable for the ORC-WER with reference and hypothesis swapped.
The basic algorithm is outlined in \cref{alg:greedy-di-cp}.
Starting with the assignment provided by the cpWER, the algorithm tests for all hypothesis segments one after the other whether changing the speaker label would improve the WER.
If so, the speaker label is modified.
The process is repeated until a stable assignment is found.
The algorithm always converges because the speaker labels are only modified when the total error improves.

\begin{algorithm}[t]
\caption{
Naive greedy approximation to DI-cpWER
}\label{alg:greedy-di-cp}
\textbf{Input}: Reference and hypothesis transcripts $\rv$, $\hv$ and speaker labels $\rlbl$, $\hlbl$ \\
\textbf{Output}: Corrected hypothesis speaker labels $\hlbl$ and the Levenshtein distance with the corrected labels
\begin{algorithmic}[1]
    \Function{LevDist}{$\rv$, $\hv$, $\rlbl$, $\hlbl$, $C_\text{S}$}
        \LComment{Compute the sum of the Levenshtein distances across all streams. Similar to \cref{eq:cp} but without the minimum operation.}
        \State
        \Return \small $\sum_{\istream'=1}^\nstream\lev\left(\select(\ind_{\{i:\hlbl_i=\istream'\}})\hv,\select(\ind_{\{i:\rlbl_i=\istream'\}})\rv, C_\text{S}\right)$
    \EndFunction
    \Function{greedyUpdateLabels}{$\rv$, $\hv$, $\rlbl$, $\hlbl$, $C_\text{S}$}   
        \While{not converged}
            \For{$\iutt\in\{1,...,\dim(\hv)\}$}
                \For{$\istream \in \{1,...,\nstream\}$}
                    \LComment{Compute Levenshtein distance with current hypothesis labels}
                    \State $d \gets$ \Call{LevDist}{$\rlbl$,$\hlbl, C_\text{S}$}
                    \LComment{Compute Levenshtein distance with updated hypothesis labels (put utterance $\iutt$ on stream $\istream$)}
                    \State $\lbl' \gets[\hlble_1,...,\hlble_{\iutt-1},c,\hlble_{\iutt + 1}, ..., \hlble_{\dim(h)}]$
                    \State $d' \gets$ \Call{LevDist}{$\rlbl$,$\lbl', C_\text{S}$}

                    \LComment{
                        Update the labels if the new distance is smaller (better)
                    }
                    \If{$d'<d$}
                        \State $\hlbl\gets\lbl'$
                    \EndIf
                \EndFor
            \EndFor
        \EndWhile
        \State \Return $\hlbl$
    \EndFunction

    \LComment{Solve \enquote{swapping} problem by running one update with a substitution cost of 2}
    \State $\hlbl \gets \Call{greedyUpdateLabels}{\rv, \hv, \rlbl, \hlbl, 2}$
    \State $\hlbl \gets \Call{greedyUpdateLabels}{\rv, \hv, \rlbl, \hlbl, 1}$

    \LComment{Return the corrected labels and Levenshtein distance}
    \State \Return $\hlbl$, \Call{LevDist}{$\rv$, $\hv$, $\rlbl$, $\hlbl, 1$}
\end{algorithmic}
\end{algorithm}

This simple algorithm often fails to solve cases where the labels of two utterances would have to be swapped to decrease the distance, but changing just one of them increases the distance.
To improve the performance in such cases, we recommend starting with a substitution cost of $C_\text{S} = C_\text{I} + C_\text{D} = 2$ so that a substitution can be traded for an insertion and a deletion until converged and then continuing with a substitution cost of 1, as indicated in the bottom of \cref{alg:greedy-di-cp}.

Note that this algorithm is not guaranteed to find the optimal solution, but it usually finds a good approximation in a reasonable amount of time (see \cref{sec:eval-greedy}).
Its complexity depends on the product of the number of output streams and reference utterances, so it is polynomial instead of exponential.
Its practical runtime depends on the speaker attribution quality.

\cref{alg:greedy-di-cp} is a naive realization and computes more than necessary:
1) Computations of the Levenshtein distance can be reused: for $d'$ only two elements of the sum in \textsc{LevDist} have to be computed. The rest is already known from the computation of $d$. 
2) Starting with the second execution of line 9, $d$ is already known; it is equal to $d$ or $d'$ from the previous iteration. 
3) Adding or removing a segment to/from a speaker stream does not require a complete re-execution of the Levenshtein algorithm.
Intermediate values from the recursive computation before the label change can be reused. 
Additionally, since the Levenshtein distance is symmetric (forward vs. reverse calculation) and the total cost can be calculated by combining forward and reverse computations, even more computations can be saved. For details, see Appendix.

\section{Error Visualization}
\label{sec:viz}

While metrics indicate the overall performance of a speech recognition model, it is impossible to tell where a system made errors from a single number.
But, this information is important when a system should be analyzed in detail and improved.
As discussed earlier, some error types cannot be reliably detected automatically, but a human can often interpret the error from a visualization.
Thus, error and alignment visualization is important for improving transcription systems.

\begin{figure}
    \tikzset{
        matchlabel/.style={circle, draw, midway,font=\scriptsize,fill=white,inner sep=.5pt},
    }
    \centering
    \begin{subfigure}[t]{0.21\textwidth}
    \centering
    \begin{tikzpicture}
        \matrix[matrix of nodes, every node/.style={inner ysep=2pt,inner xsep=1.5pt}] (m) {
            I & N & D & U & S & T & * & R & Y & * & * \\[.5em]
            I & N & * & * & * & T & E & R & E & S & T \\
        };
    \end{tikzpicture}
    \subcaption{\emph{Alignment:} \emph{nulls} (*) are inserted so that both sequences have the same length and matching tokens align.
    }\label{fig:align-alignment}
    \end{subfigure}
    \hspace*{1em}
    \begin{subfigure}[t]{0.21\textwidth}
    \centering
    \begin{tikzpicture}
        \matrix[matrix of nodes, every node/.style={inner ysep=2pt}] (m) {
            I & N & D & U & S & T & R & Y \\[1.5em]
            I & N & T & E & R & E & S & T \\
        };
        \draw[black] 
            (m-1-1.south) --node[matchlabel,fill=correct]{C} (m-2-1.north) 
            (m-1-2.south) --node[matchlabel,fill=correct]{C} (m-2-2.north) 
            (m-1-6.south) --node[matchlabel,fill=correct]{C} (m-2-3.north)
            (m-1-7.south) --node[matchlabel,fill=correct]{C} (m-2-5.north)
            (m-1-8.south) --node[matchlabel,fill=substitution]{S} (m-2-6.north);
    \end{tikzpicture}
    \subcaption{\emph{Trace:} Matching tokens are connected with a line. Correct or Substitution is marked with \markC{} and \markS.} \label{fig:align-trace}
    \end{subfigure}
    \caption{Different ways for visualizing a Levenshtein matching defined in \protect\cite{kruskal1983overview}. Note that multiple alignments can correspond to the same trace.}
    \label{fig:viz-align}
\end{figure}

\tikzset{
        matchlabel/.style={circle, draw, midway,font=\scriptsize,fill=white,inner sep=.5pt},
    }

Sequence alignments can be visualized in different ways, as shown in \cref{fig:viz-align} \cite{kruskal1983overview}.
The plot uses characters instead of words to shorten the visualizations.
The \emph{alignment} visualization in \cref{fig:align-alignment} inserts null tokens (here: *) into both sequences so that their lengths match and matching words have the same positions.
This visualization is ideal for single-sentence alignments without temporal information because the matching appears relatively clean.
It can become cluttered and hard to grasp for longer sequences and with more than a single speaker.

In the \emph{trace} visualization in \cref{fig:align-trace}, matching words are connected with lines.
Insertions and deletions are not connected, so any unmatched words in the reference or hypothesis are deletions and insertions, respectively.
This allows for retaining the temporal annotations in the visualization by positioning words according to their timestamps.
Regions with no speech activity are empty and errors can directly be related to their temporal position.
Combined with colorization, this leads to a visualization that is easy to grasp and interpret.
It works for long sequences with many speakers.

Existing tools for alignment visualization often use the\emph{alignment} style and are thus not applicable to meeting scenarios.
The Kaldi toolkit, for example, provides a text-based visualization which breaks apart when the sequences span more than one line in a text editor.
Alignment tools from Biology typically focus on multiple-sequence alignment.
No tool that we know of considers temporal alignment and timestamps.
We found that it can be immensely helpful to precisely identify (temporal) positions where the system made errors, to be able to listen to the input signals, and to examine the system's internals in detail.
For example, in \cref{fig:wer-visualization}, one can deduce from the difference between cpWER and DI-cpWER that there is at least one speaker attribution error.
But for improving the system, it is beneficial to know that there is a speaker counting error and that the words \enquote{C}, \enquote{D}, \enquote{E} and \enquote{H} got assigned a wrong speaker label.

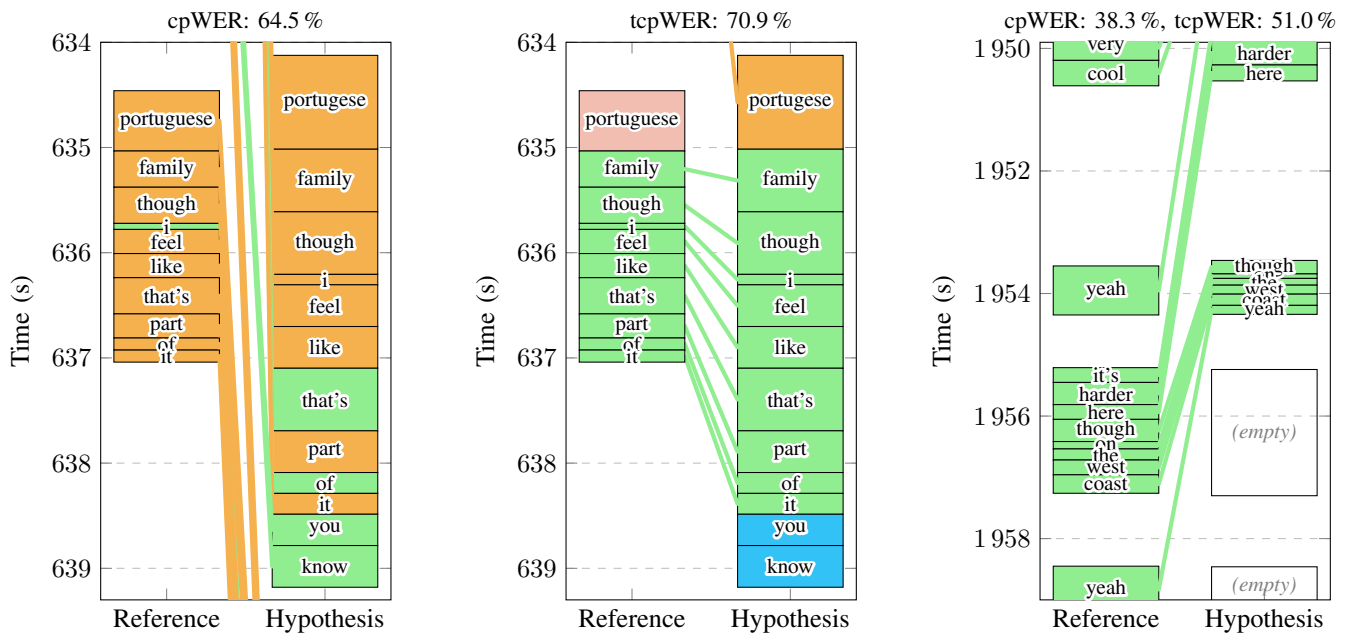
\begin{figure*}[b]
    \centering
    \begin{subfigure}[t]{0.3\textwidth}
        \centering%
        \input{plots/ex3.tex}%
        \caption{Implausible matching produced by a non-time-constrained WER, such as cpWER or ORC-WER. The transcript is correct but words are matched as substitutions across large temporal distances.}
        \label{fig:cp-tcp-ex3}
    \end{subfigure}
    \hspace*{5mm}
    \begin{subfigure}[t]{0.3\textwidth}
        \centering%
        \input{plots/ex1.tex}%
        \caption{The same example as \cref{fig:cp-tcp-ex3}, but with a time-constrained WER. The matching is now more plausible. \enquote{portugese} is not matched with \enquote{portuguese} because the matching is ambiguous.}
        \label{fig:cp-tcp-ex1}
    \end{subfigure}
    \hspace*{5mm}
    \begin{subfigure}[t]{0.3\textwidth}
        \centering%
        \input{plots/ex2.tex}%
        \caption{cpWER matching of decoupled diarization and recognition: empty segments may result in substitutions instead of deletions. tcpWER would count these as deletions.}
        \label{fig:cp-tcp-ex2}
    \end{subfigure}

    \caption{Trace visualizations excerpts from a single speaker from the DiPCo corpus. 
    Time flows from top to bottom.
    Boxes represent word start and end times and matched words are connected with a line.
    Connecting lines may connect words that lie outside of the displayed excerpt.
    Matchings are color-coded as correct \legend{correct}, substitution \legend{substitution}, deletion \legend{deletion}, and insertion \legend{insertion}.
    }
    \label{fig:examples-cp-tcp}
\end{figure*}

The MeetEval toolkit (see \cref{sec:open-source}) provides tools for generating interactive \emph{trace} visualizations.
Examples are available online\footnote{\url{https://fgnt.github.io/meeteval_viz}} and also displayed as simplified versions in \cref{fig:examples-cp-tcp}.

\section{Experiments and Discussion}

The experiments aim to show relations between metrics and their responses to different error types found in real meetings.

\subsection{Data and systems}
\label{sec:data}
We mainly use the submission results from the CHiME-7 challenge \cite{Cornell2023_CHiME7DASRChallenge} for our experiments and kindly thank the participants for permitting us to use their submissions \cite{wang2023ustcnercslip,kamo2023multispeaker,prisyach2023stcon,deng2023university,boeddeker2023multistage}.
By doing so, we collected results from a variety of systems to show that the metrics are not biased towards a specific system.
Conclusions drawn from the experiments generalize across different architectures.
We explicitly exclude the results on the Mixer6 dataset due to its license restrictions.

We additionally perform some experiments on the LibriCSS corpus \cite{Chen2020_ContinuousSpeechSeparation}.
LibriCSS contains re-recorded meetings constructed from LibriSpeech \cite{panayotov2015librispeech}.
Five 1h-long sessions have been recorded with different overlap ratios, which we call LibriCSS-1h.
These sessions have been sub-divided two times, into 10-minute-long recordings (LibriCSS-raw) and into segments of one to two minutes in length (LibriCSS-segments).
These different sub-segmentations allow for analysis of the impact of temporal length on the metrics while keeping the speech content the same.
For experiments on LibriCSS, we use the \gls{CSS} and ASR pipeline from \cite{neumann2023meeting}.

\subsection{Time-constrained word error rates}
\label{sec:eval-time-constrained}
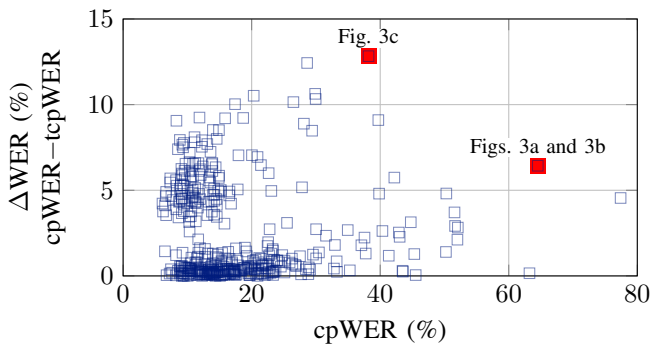
\begin{figure}
    \input{plots/cp-vs-tcp.tex}
    \caption{Scatter plot of tcpWER vs cpWER for examples across datasets. Their difference shows implausible matchings and large differences are a hint for severe temporal errors. The two highlighted examples are shown in \cref{fig:examples-cp-tcp} and discussed in detail in \cref{sec:eval-time-constrained}. The WERs differ by less than 10 percentage points for most examples.}
    \label{fig:cp-vs-tcp}
\end{figure}

\cref*{fig:cp-vs-tcp} shows a scatter plot comparing tcpWER with cpWER across all submissions and datasets.
It is immediately visible that the cpWER provides a lower bound on the tcpWER, i.e., $\text{tcpWER}-\text{cpWER}\geq 0$.
It can also be seen that the difference between the two metrics is usually below 10 percentage points.
This difference can be regarded as significant, especially in the low-WER regions where system performance often only differs by less than a percentage point.
cpWER and tcpWER are correlated for our collection of experiments.

Two interesting examples are visualized as timelines, where time flows from top to bottom, in \cref{fig:examples-cp-tcp}.
\cref{fig:cp-tcp-ex3} and \cref{fig:cp-tcp-ex1} show the same example matched with tcpWER and cpWER, respectively.
Most words are transcribed correctly and at the correct temporal positions, but the cpWER matches them with substitutions with temporally far away words, which can be seen by the orange connecting lines that leave the displayed excerpt in \cref{fig:cp-tcp-ex3}.
This happens when the number of errors at another temporal position can be reduced by matching a far away word as a substitution.
This example shows that non-time-constrained matchings can be implausible and that a time constraint is required for realistic matchings.
This is directly reflected in the metric values: the cpWER is significantly lower than the tcpWER.
This is the main cause for differences between tcpWER and cpWER in our collection of systems.

In \cref{fig:cp-tcp-ex2}, transcription is decoupled from diarization, i.e., both words and speech activity are estimated (roughly) correctly, but the words are placed in the wrong segments.
This error is again reflected in the tcpWER of \SI{51.0}{\percent}, but not in the cpWER of \SI{38.2}{\percent}, which confirms that the tcpWER increases with poor temporal alignments.
The difference between cpWER and tcpWER can be used to detect such cases where the mapping of words to speech segments is incorrect.

The issues shown in \cref*{fig:examples-cp-tcp} are impossible to detect by a metric that evaluates transcripts or diarization in isolation.
Even the DA-WER does not capture these issues because, although it evaluates both diarization and speech recognition, it evaluates them only loosely coupled.
We conclude here that a metric for long-form meeting recognition should evaluate diarization and speech recognition jointly, such as the tcpWER, to get a realistic impression of the performance.
Also, a visualization as proposed in \cref{sec:viz} is important to find issues that are not (directly) reflected in the metric or cannot be discriminated from the metric.

\subsection{ORC-WER and MIMO-WER}

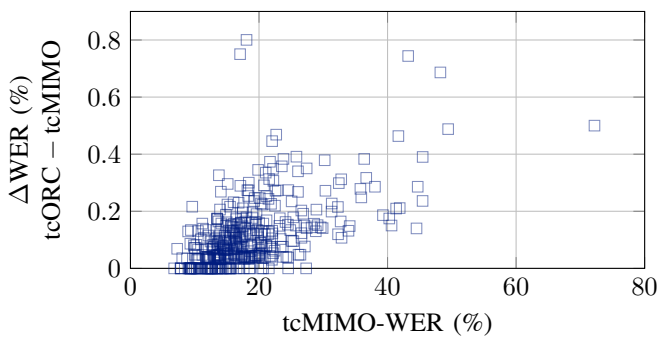
\begin{figure}
    \input{plots/orc-vs-mimo.tex}%
    \caption{
        Scatter plot of tcMIMO-WER vs tcORC-WER for examples across datasets. 
        Their difference is smaller than 0.2 percentage points on average on this data. The ORC-WER is thus a good replacement for the MIMO-WER for this set of data and systems.
    }
    \label{fig:orc-vs-mimo}
\end{figure}

tcORC-WER and tcMIMO-WER are compared in \cref{fig:orc-vs-mimo}.
We here compare the time-constrained versions because the non-time-constrained variants are not feasible to compute on this data.
The two metrics are strongly correlated and differ by less than 0.2 percentage points on average.
This reflects the discussion in \cref{sec:mimo} that major differences only occur when a system disrespects the physical utterance order.

Minor differences stem from edge cases where the order of segments is unclear, e.g., when two segments have a similar begin time, or when poor separation or similar words lead to ambiguous assignments.
This experiment shows that in most cases, ORC-WER or tcORC-WER can be used as a good proxy for the MIMO-WER or tcMIMO-WER with a lower computational complexity.

\subsection{Behavior under timestamp errors}

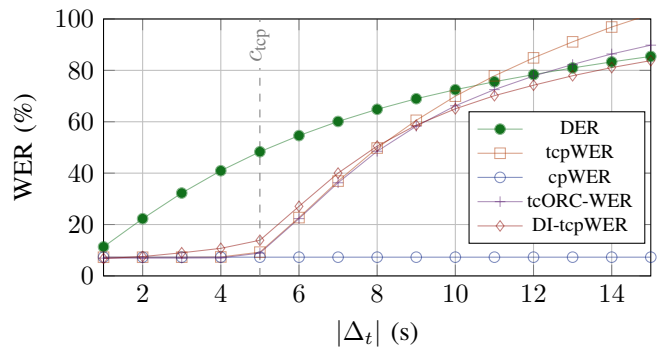
\begin{figure}
    \input{plots/timestamp-precision.tex}
    \caption{Influence of a jitter $|\Delta_t|$ in the timestamps on different metrics. The cpWER is not influenced by the jitter, while the tcpWER and DER are.}
    \label{fig:timestamp-precision}
\end{figure}

\cref{fig:timestamp-precision} shows the influence of a jitter in the timestamps on different metrics.
The experiments were conducted on the LibriCSS-raw dataset by adding random noise to the timestamps, drawn uniformly from the interval $[-\Delta_t, \Delta_t]$.
The cpWER is independent of the timestamps and thus constant across different jitters.
Both tcpWER and DER increase with increasing jitter with comparable behavior except for an offset in the tcpWER caused by the collar of \SI{5}{s}.
A jitter smaller than the collar does not affect the tcpWER, but any jitter larger than that increases the value significantly.

\subsection{Behavior under speaker label errors}

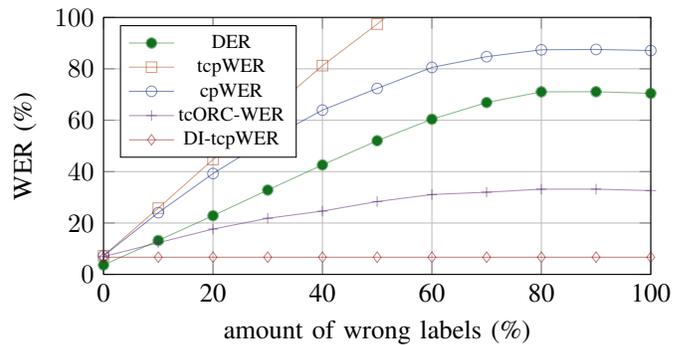
\begin{figure}
    \input{plots/label-switches.tex}%
    \caption{Influence of wrongly assigned speaker labels on different WERs on the LibriCSS dataset. Speaker label errors are simulated by randomly assigning speakers to segments in the hypothesis.}
    \label{fig:label-switches}
\end{figure}

The influence of wrongly assigned speaker labels is analyzed in \cref{fig:label-switches} by artificially swapping speaker labels in the hypothesis, again on the LibriCSS-raw dataset.
The value of the metrics is shown over the percentage of randomly swapped hypothesis labels.
While both ORC-WER and DI-cpWER are somewhat agnostic to diarization errors, only the DI-cpWER is fully invariant to label swaps on the hypothesis side.
The ORC-WER increases with the percentage of swapped labels because it cannot correct utterance split errors (see \cref{sec:error-types}).
The non diarization-invariant metrics tcpWER and DER increase similarly with an increasing number of wrong speaker labels.

\subsection{Greedy Approximation to DI-cpWER}
\label{sec:eval-greedy}

\begin{figure}
    \input{plots/ditcp-vs-invtcorc.tex}%
    \caption{Scatter plot comparing the exact and greedy algorithms for the DI-tcpWER for examples across datasets.  The greedy algorithm finds the correct assignment 86\% of time, and its results differ from the exact computation by less than 0.02 percentage points on average.}
    \label{fig:ditcp-vs-invtcorc}
\end{figure}
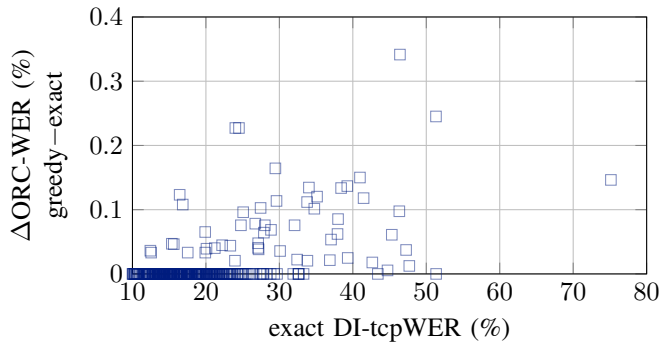

\cref*{fig:ditcp-vs-invtcorc} compares the DI-cpWER computed with the greedy algorithm with the exact computation \footnote{Note that these results also hold for the ORC-WER which uses the same alignment algorithms}.
The plot barely shows a difference between the two algorithms on this data.
The greedy algorithm finds the exact matching 86\% of time and the average difference is less than 0.02 percentage points.
This shows that the greedy algorithm approximates the DI-tcpWER with a tolerance that is typically below statistical significance.

\subsection{System analysis with metrics combinations}

Some errors described in \cref{sec:error-types} cannot be detected by a single metric, such as the number of utterance splits and merges.
The impact of these errors, however, can be estimated by comparing different metrics.

A low DI-cpWER combined with a high ORC-WER or MIMO-WER indicates utterance split errors while a low ORC-WER combined with a high DI-cpWER indicates utterance merge errors.
If DI-cpWER and ORC-WER are similar, utterance merges and splits are likely balanced.
As already shown in \cref{sec:eval-time-constrained}, a high time-constrained error rate (tcpWER, tcORC-WER or tcMIMO-WER) combined with a corresponding low non-time-constrained error rate (cpWER, ORC-WER or MIMO-WER) indicates timing issues.

A low (tc)DI-cpWER combined with a high (t)cpWER indicates speaker label errors, since (tc)DI-cpWER is invariant to speaker label switches on the hypothesis side.
As an example, in \cite{Boeddeker2024_OnceMoreDiarization} Table 1 and \cite{neumann2023meeting} Table 2, the DI-cpWER was used as an oracle to analyze the potential performance (upper bound) that can be achieved by only improving the speaker attribution.
In both papers, the DI-cpWER yielded a valuable insight into the errors and helped to identify that an improved diarization could yield a significant improvement in the overall system performance.
Modifications to the systems, which were made in reaction to the gap between cpWER and DI-cpWER, reduced the gap, showing the the modifications successfully improved the speaker attribution estimation.

\subsection{Runtime performance}
\label{sec:benchmark}

\begin{figure}
    \input{plots/benchmark.tex}%
    \caption{Execution time over signal length for time-constrained vs non-time-constrained ORC-WER and MIMO-WER. The metrics were computed for the output of a 2-output CSS system on LibriCSS. Triangles on the top margin indicate infeasible examples (exceeded 20 minutes of execution time).}
    \label{fig:benchmark}
\end{figure}
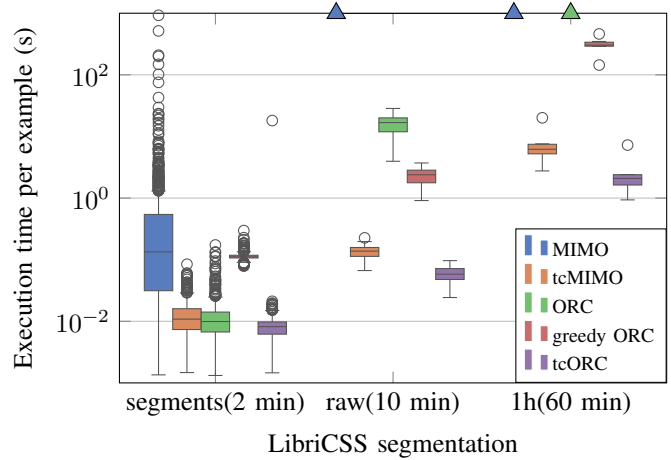

The time constraint and the greedy algorithm introduced in this work significantly improve the execution time.
This is shown in \cref{fig:benchmark} on the different sub-sets of LibriCSS using a \gls{CSS}-style separator with two output channels.
All algorithms are implemented in C++ except for the greedy algorithm, which is implemented in Python using numpy, i.e., a mix of a scripting language and C++.
All metrics can be computed on the short LibriCSS-segments dataset, but as the length increases, ORC-WER and MIMO-WER quickly become infeasible.
The outliers for MIMO-WER indicate that the execution time can become large even for short examples.
The time constraint significantly improves the execution time so that the runtime is practically negligible on average even for the 1h dataset.
The greedy algorithm allows for computing the non-time-constrained ORC-WER for the long examples.
It is natural for the greedy implementation to be slower than the exact version on the segments dataset because it is implemented with a slower programming language.

\section{Open Source}
\label{sec:open-source}
All algorithms presented in this work are available in the open-source toolkit MeetEval at \url{https://github.com/fgnt/meeteval}.
The results in this work were created with version \texttt{0.4.1}.
The visualization tool for generating the visualizations discussed in \cref{sec:viz} is available as a command line tool in MeetEval, but can also be executed in the browser at \url{https://fgnt.github.io/meeteval_viz}.
Simplified implementations and descriptions of the algorithms are available at \url{https://github.com/fgnt/meeteval/blob/main/doc/algorithms.md}.

\section{Summary}

In this work, we presented a definition for computing WERs for meeting-level speech recognition that unifies many definitions from the literature.
We highlighted their advantages and disadvantages and found that different WER definitions highlight different error types, but it is impossible to disentangle their influence on the final metric value.
To at least get an estimate for the number of errors caused by speaker assignment errors, we proposed the DI-cpWER and showed that it is invariant to speaker label switches.
We also incorporated the time constraint from the previously presented tcpWER into ORC-WER and MIMO-WER.
We showed that the resulting tcORC-WER and tcMIMO-WER provide more plausible matching and are faster to compute compared to their non-time-constrained counterparts.
For cases where the DI-cpWER or ORC-WER are not feasible, we proposed a greedy algorithm to approximate their value with high accuracy and reduced computational complexity.
We finally presented a way of visualizing the sequence alignment between reference and hypothesis transcripts for detailed system analysis to facilitate spotting of errors in the system output.

\bibliographystyle{IEEEtran}
\bibliography{references}

\appendices
\section*{Efficient Implementation of the Greedy Algorithm for DI-cpWER and ORC-WER}
\label{sec:appendix-greedy}

The naive greedy algorithm from \cref{alg:greedy-di-cp} can be optimized by re-using previously computed states.

Let $m_{\istream\iutt}=\lev(\select(\ind_{\{i:\hlble_i = c\} \cup \{\iutt\}}\hv, ...))$ be the Levenshtein distance of stream $\istream$ with utterance $\iutt$ placed on that stream and $n_{\istream\iutt}=\lev(\select(\ind_{\{i:\hlble_i = c\} \setminus \{\iutt\}}\hv, ...))$ be the Levenshtein distance of stream $\istream$ with utterance $\iutt$ removed (if it was present).
The update of $\hlble_\iutt$ from \cref{alg:greedy-di-cp} (inner for loop) can then be rewritten as
\begin{align}
    \hlble_\iutt &= \argmin_{1\leq\istream\leq\nstream} m_{\istream\iutt} + \sum_{\istream'\neq\istream} n_{\istream'\iutt} \label{eq:greedy1}\\
    &= \argmin_{1\leq\istream\leq\nstream} m_{\istream\iutt} - n_{\istream\iutt}, \label{eq:greedy2}
\end{align}
where one of $m_{\istream\iutt}$ or $n_{\istream\iutt}$ is known from an earlier iteration.
\cref{eq:greedy2} follows from \cref{eq:greedy1} by pulling the sum $\sum_{\istream'=1}^\nstream n_{\istream'\iutt}$ out of the minimum.
Using this update instead of the loop in \cref{alg:greedy-di-cp} reduces the number of times that $\lev$ is applied for every utterance and iteration from $2\nstream^2$ to $\nstream$.

The remaining Levenshtein distance (re-)computations can further be sped up by re-using parts of the Levenshtein matrix.
We arrange the intermediate results of the Wagner-Fisher algorithm from \cref{eq:lev} in a matrix $\matr L = [l_{rh}]_{rh}=[\lev(\select(\vect v_r)\rv,\select(\vect v_h)\hv)]_{rh}$, where $\vect v_i = [1,...,1,0,...,0]\tran\in\{0,1\}^\nstream$ is the vector that contains exactly $i$ ones and is padded at the end with zeros up to length $\nstream$.

The next optimization follows from the observation that the Levenshtein distance is symmetric, i.e., $\lev(\rv,\hv)=\lev(\reversed(\rv), \reversed(\hv))$, where $\reversed$ reverses the sequence.
In fact, if we denote with $\matr L^\text{(fw)}$ the Levenshtein matrix for $\lev(\rv,\hv)$ and with $\matr L^\text{(rev)}$ the Levenshtein matrix for $\lev(\reversed(\rv), \reversed(\hv))$, we can find the total Levenshtein distance as the minimum over any column of $\matr L^\text{(comb)} = \matr L^\text{(fw)} + \reversed(\matr L^\text{(rev)})$, where $\reversed$ reverses the order of the columns of $\matr L^\text{(rev)}$: $\forall h: \lev(\rv,\hv) = \min_r l^\text{(comb)}_{r,h}$.
This property can be seen when realizing that $l^\text{(comb)}_{rh} = l_{rh}^\text{(fw)} + l_{\dim(\rv) - r, \dim(\hv) - h}^\text{(rev)}$ is the Levenshtein distance between $\rv$ and $\hv$ assuming that the $r$-th element from $\rv$ is matched with the $h$-th element from $\hv$.
The minimum over $r$ is thus the Levenshtein distance assuming that the $h$-th element from $\hv$ is matched with any element from $\rv$, which is the total Levenshtein distance.

To find $m_{cu}$, we can now use the column from $\matr L^\text{(fw)}$ that corresponds to the position where the utterance $u$ should be placed and compute the missing columns for $u$ with \cref{eq:lev}.
We then add the last newly computed column of $\matr L^\text{(fw)}$ to the corresponding column in $\matr L^\text{(rev)}$ and compute the minimum to obtain $m_{cu}$.
The Levenshtein distance for a removed segment $n_{cu}$ can be found similarly by adding the column from $\matr L^\text{(fw)}$ right before utterance $u$ to the column from $\matr L^\text{(rev)}$ right behind utterance $u$ and computing the minimum.

This optimization speeds up the computation of $m_{cu}$ and $n_{cu}$ significantly, especially when the reference and hypothesis contain many segments.
Computing $m_{cu}$ or $n_{cu}$ only needs as much time as computing the Levenshtein distance for a single segment instead of the full stream, at the expense of pre-computing the reverse Levenshtein matrix $\matr L^\text{(rev)}$ once.

Additionally, the matrices do not have to be fully stored in memory.
The reverse matrix $\matr L^\text{(rev)}$ has to be pre-computed, but unused columns can be discarded when progressing forward through the utterances.
The forward matrix $\matr L^\text{(fw)}$ can be computed on-the-fly and only the last column (right before the position of $u$) has to be stored in memory for every stream.

A detailed outline of this algorithm can be found as supplementary material\footnote{\url{https://github.com/fgnt/meeteval/blob/main/doc/algorithms.md\#faster-version-of-the-greedy-orc-levenshtein-distance-algorithm}}.

\end{document}

%% file: macros.tex
\usepackage{amsmath}
\usepackage{amssymb}
\usepackage{mathabx}
\usepackage{mathtools}
\usepackage{bbm}
\usepackage{physics}
\usepackage{siunitx}
\usepackage{trfsigns}
\usepackage{pifont}
\usepackage{etoolbox}
\usepackage{multirow}
\usepackage{csquotes}
\usepackage[shortlabels]{enumitem}  %
\usepackage{cite}   %
\usepackage{hyperref}
\usepackage[capitalize]{cleveref}
\usepackage{xifthen}
\usepackage{balance} %
\usepackage{booktabs}
\usepackage{listings}
\usepackage{listofitems}
\usepackage{tabularx}
\usepackage{varwidth}
\usepackage{cprotect}  %
\usepackage{sepfootnotes} %

\usepackage{url}

\usepackage{algorithm}
\usepackage[]{algpseudocodex}
\algtext*{EndWhile}%
\algtext*{EndIf}%
\algtext*{EndIf}%

\algnewcommand{\GoTo}[1]{\textbf{go to}~\ref{#1}}%

\newcolumntype{H}{>{\setbox0=\hbox\bgroup}c<{\egroup}@{}}   %
\newcommand{\tbl}[2][c]{\begin{tabular}{@{}#1@{}}#2\end{tabular}}   %

\newcommand*{\thl}{\fontseries{b}\selectfont}
\robustify\thl
\robustify\fontseries

\newcommand{\inred}[1]{\textcolor{red}{#1}}

\usepackage{xcolor}
\usepackage[normalem]{ulem} %
\definecolor{diffaddfg}{HTML}{54aeff}
\definecolor{diffaddbg}{HTML}{ddf4ff}
\definecolor{diffdelfg}{HTML}{ffb77c}
\definecolor{diffdelbg}{HTML}{fff1e5}

\ifdefined\annotated
\definecolor{diffaddfg}{HTML}{007FFF} %
\definecolor{diffdelfg}{HTML}{FF4500} %
\newcommand\hladd{%
  \bgroup
  \markoverwith{\textcolor{diffaddbg}{\rule[-.5ex]{.1pt}{2.5ex}}}%
  \ULon}
\newcommand\hlsout{%
  \bgroup
  \markoverwith{\textcolor{diffdelbg}{\rule[-.5ex]{.1pt}{2.5ex}}\llap{\rule[.5ex]{.3pt}{0.6pt}}}%
  \ULon}
\newcommand{\added}[1]{%
    \hladd{#1}%
}
\newcommand{\removed}[1]{\hlsout{#1}}
\newcommand{\modified}[2]{\hlsout{#1}\hladd{#2}}
\newcommand{\removednohl}[1]{#1}

\robustify{\added}
\robustify{\removed}
\robustify{\modified}

\newcommand\mremoved[1]{{%
    \savebox{0}{\ensuremath{#1}}\mathrlap{%
    \smash{\textcolor{diffdelbg}{\rule[-.25ex]{\wd0}{\ht0}}}}{\ensuremath{#1}}\mathllap{\rule[.5ex]{\wd0}{0.6pt}}}}

\else
\newcommand{\added}[1]{#1}
\newcommand{\removed}[1]{}
\newcommand{\removednohl}[1]{}
\newcommand{\modified}[2]{#2}

\newcommand{\mremoved}[1]{}
\fi

\let\oldcite\cite
\renewcommand{\cite}[1]{%
\ifthenelse{\isempty{#1}}%
{\inred{[cite!]}}%
{\unskip\mbox{\oldcite{#1}}}%
}

%% file: tikz_style.tex
\usepackage{tikz}
\usepackage{pgfplots}
\pgfplotsset{compat=1.16}
\usepgfplotslibrary{groupplots}
\usetikzlibrary{positioning,arrows,matrix,fit,calc,patterns,chains,scopes,shapes.multipart,decorations,decorations.markings,backgrounds,math,arrows.meta,pgfplots.statistics, pgfplots.colorbrewer}

\definecolor{palette-1}{HTML}{001C7F}    
\definecolor{palette-2}{HTML}{B1400D}    
\definecolor{palette-3}{HTML}{12711C}    
\definecolor{palette-4}{HTML}{8C0800}    
\definecolor{palette-5}{HTML}{591E71}    
\definecolor{palette-6}{HTML}{592F0D}    
\definecolor{palette-7}{HTML}{A23582}    
\definecolor{palette-8}{HTML}{3C3C3C}    
\definecolor{palette-9}{HTML}{B8850A}    
\definecolor{palette-10}{HTML}{006374}      

\definecolor{insertion}{HTML}{33c2f5}
\definecolor{deletion}{HTML}{f2beb1}
\definecolor{substitution}{HTML}{F5B14D}
\definecolor{correct}{HTML}{90ee90}

\tikzset{
    line/.style={draw,black,thick,rounded corners=1mm,line cap=round},
    noshortarrow/.style={line,->},
    arrow/.style={noshortarrow,shorten >=.3mm},
    doublearrow/.style={arrow,<->, shorten <=.3mm},
    box/.style={draw,black,thick,minimum height=3em,text height=1.5ex,text depth=0.25ex,rounded corners=3,fill=white},
    nopadding/.style={minimum height=0,inner sep=1mm},
    signalbox/.style={draw,black,thin,rounded corners=1mm,minimum width=7mm, minimum height=4mm,inner sep=0},
    pbox/.style={box,fill=black!10},
    backgroundbox/.style={inner xsep=3mm, inner ysep=1mm, draw, dashed, rounded corners,fill=orange!10},
    branch/.style={inner sep=0.3mm,circle,fill=black},
    operator/.style={draw,circle,black,rounded corners,inner sep=0,fill=white},
    vertex/.style={draw,ultra thin,circle,black,rounded corners,inner sep=0.6mm,fill=gray,fill opacity=0.5},
    edge/.style={line,very thick,line cap=butt},
    pattern1/.style={pattern=north west lines,pattern color=palette-1},
    pattern2/.style={pattern=north east lines,pattern color=palette-2},
    pattern3/.style={pattern=crosshatch,pattern color=palette-3},
    buswidth/.style={path picture={\draw[black,-] (path picture bounding box.south west) -- (path picture bounding box.north east);}}
}

\tikzset{
    matchlabel/.style={circle, draw, midway,font=\small,fill=white,inner sep=.5pt,text=black},
}

\DeclareRobustCommand\markC{\tikz[baseline=-.6ex]{ \node[matchlabel,fill=correct]{C};}}
\DeclareRobustCommand\markS{\tikz[baseline=-.6ex]{\node[matchlabel,fill=substitution]{S};}}
\DeclareRobustCommand\legend[1]{\tikz[baseline=-.6ex]{\node[fill=#1,draw=black,minimum width=1ex,minimum height=1ex]{};}}

\pgfdeclarelayer{bg}
\pgfdeclarelayer{bbg}
\pgfdeclarelayer{fg}
\pgfdeclarelayer{ffg}
\pgfsetlayers{bbg,bg,main,fg,ffg}

\tikzset{
    viztext/.style={font=\footnotesize},
    viztext-empty/.style={font=\footnotesize\itshape,gray,pos=.5},
    match/.style={ultra thick,line cap=round},
    match-substitution/.style={match,draw=substitution},
    match-insertion/.style={match,draw=insertion},
    match-deletion/.style={match,draw=deletion},
    match-correct/.style={match,draw=correct},
}

%% file: symbols.tex
\newcommand\defm[2]{\expandafter\newcommand{#1}{\ensuremath{#2}}} %

\defm\Loss{\mathcal{L}}    %

\newcommand{\matr}[1]{\ensuremath{\boldsymbol{\mathbf{#1}}}}
\newcommand{\vect}[1]{\ensuremath{\boldsymbol{\mathbf{#1}}}}
\newcommand*{\tran}{^{\mkern-1.5mu\mathsf{T}}}

\DeclareMathOperator*{\argmin}{arg\,min}

\defm{\lev}{\mathrm{lev}}
\DeclareMathOperator*{\tclev}{tclev}

\defm{\rv}{\vect{r}}
\defm{\hv}{\vect{h}}
\defm{\lble}{\ell}
\defm{\lbl}{\vect{\lble}}
\defm{\rlbl}{\lbl^\text{ref}}
\defm{\hlbl}{\lbl^\text{hyp}}
\defm{\rlble}{\lble^\text{ref}}
\defm{\hlble}{\lble^\text{hyp}}
\defm{\trne}{t} %
\defm{\trn}{\vect{\trne}}    %
\defm{\rtrn}{\trn^\text{ref}}
\defm{\htrn}{\trn^\text{hyp}}
\defm{\rtrne}{\trne^\text{ref}}
\defm{\htrne}{\trne^\text{hyp}}
\defm{\tbege}{b}  %
\defm{\tbeg}{\vect{\tbege}}  
\defm{\rtbeg}{\tbeg^\text{ref}}
\defm{\htbeg}{\tbeg^\text{hyp}}
\defm{\rtbege}{\tbege^\text{ref}}
\defm{\htbege}{\tbege^\text{hyp}}
\defm{\tende}{e}  %
\defm{\tend}{\vect{\tende}}  
\defm{\rtend}{\tend^\text{ref}}
\defm{\htend}{\tend^\text{hyp}}
\defm{\rtende}{\tende^\text{ref}}
\defm{\htende}{\tende^\text{hyp}}
\DeclareMathOperator*{\select}{sel}
\defm{\selection}{\matr{S}}

\defm{\tref}{\mathcal{R}} %
\defm{\sref}{r} %
\defm{\thyp}{\mathcal{H}}   %
\defm{\shyp}{h} %
\defm{\nsref}{N^\text{ref}}  %
\defm{\nshyp}{N^\text{hyp}}  %

\defm{\idx}{p}  %
\defm{\ridx}{p_\text{ref}}
\defm{\hidx}{p_\text{hyp}}
\defm{\seg}{s}  %
\defm{\ind}{\mathbbm{1}}    %

\DeclareMathOperator{\distance}{\lev}
\DeclareMathOperator{\reversed}{rev}
\defm{\nspk}{K}
\defm{\ispk}{k}
\defm{\nstream}{C}
\defm{\istream}{c}
\defm{\nutt}{U}
\defm{\iutt}{u}
\defm{\perm}{\pi}
\defm{\permset}{\mathcal{P}}
\defm{\assignment}{\vect{a}}
\defm{\collar}{c}

%% file: glossaries.tex
\usepackage[acronym,shortcuts]{glossaries}
\glsdisablehyper    %

\newacronym{cpWER}{cpWER}{Concatenated minimum-Permutation Word Error Rate}
\newacronym{DA-WER}{DA-WER}{Diarization-Aware Word Error Rate}
\newacronym{DER}{DER}{Diarization Error Rate}
\newacronym{WER}{WER}{Word Error Rate}
\newacronym{ORC-WER}{ORC-WER}{Optimal Reference Combination Word Error Rate}
\newacronym{tcORC-WER}{tcORC-WER}{Time-Constrained ORC-WER}
\newacronym{MIMO-WER}{MIMO-WER}{Multiple Input Multiple Output Word Error Rate}
\newacronym{tcMIMO-WER}{MIMO-WER}{time-constrained MIMO-WER}
\newacronym{tcpWER}{tcpWER}{Time-Constrained minimum-Permutation Word Error Rate}
\newacronym{CSS}{CSS}{Continuous Speech Separation}
\newacronym{DI-cpWER}{DI-cpWER}{Diarization-Invariant cpWER}
\newacronym{DI-tcpWER}{DI-tcpWER}{Diarization-Invariant tcpWER}
\newacronym{CER}{CER}{Character Error Rate}
\newacronym{SER}{SER}{Sentence Error Rate}
\newacronym{SOT}{SOT}{Serialized Output Training}

%% file: wer-definition-figure.tex
\newcommand{\tabitem}{~~\llap{\textbullet}~}
\newsavebox\mathbox
\newcommand{\cbox}[2]{%
    \savebox{\mathbox}{\ensuremath{#2}}\mathrlap{%
    \smash{\textcolor{#1}{\hspace{-.05em}\rule[-.25ex]{\wd\mathbox}{\ht\mathbox}}}}{\ensuremath{#2}}}

\newcommand\segm[4]{
    \begin{scope}[local bounding box=bb]
        \foreach \word/\xpos/\color in {#2} {
            \node[w] at (\xpos|-#1) (#3-\word) {\word};
        }
        \begin{pgfonlayer}{bg}
            \node[u,fit={(bb)},fill=white,draw=none](#3-#4){};
            \begin{scope}
                \clip[rounded corners=2] (#3-#4.north west) rectangle (#3-#4.south east);
                \foreach \word/\xpos/\color in {#2} {
                    \ifthenelse{\equal{\xpos}{\color}}{
                    }{
                        \path[fill=\color,line width=0] ($(#3-\word.north west) + (-.1em,1)$) rectangle ($(#3-\word.south east) + (.1em,-1)$);
                    }
                }
            \end{scope}
            \draw[u] (#3-#4.north west) rectangle (#3-#4.south east);
        \end{pgfonlayer}
    \end{scope}
}

\tikzset{
    u/.style={draw=black,inner sep=.1em, outer sep=0, rounded corners=2,anchor=west},
    w/.style={font=\footnotesize,inner xsep=.5,inner ysep=.5,outer sep=0,rounded corners=2,anchor=center},
    uc/.style={start chain=chain,node distance=0,anchor=Center},
    label/.style={anchor=east,inner sep=1},
    m/.style={matrix of nodes,every node/.style={anchor=east}},
    assign/.style={-{Latex[length=.2em,width=.3em]},rounded corners=1.5,thin},
    arrow/.style={->,thin},
    shadow/.style={draw=red!20,color=red!20},
    matchcorr/.style={draw=green,line cap=round},
    matchsub/.style={draw=red,line cap=round},
    transcriptbox/.style={fill=black!10,draw=black,inner sep=.2em,inner xsep=.5em, rounded corners=2},
    matchingbox/.style={transcriptbox,fill=palette-1!30},
    pics/utt/.style args={#1,#2}{
        code={
            \begin{scope}[start chain=chain,node distance=0,local bounding box=bb]
                \foreach \c/\d in {#1} {
                    \node[w,on chain](#2\c){\phantom{\c}};
                };
                \node[u,fit={(bb)},draw=none](#2){};
                \path[clip,rounded corners=2] (#2.north west) rectangle (#2.south east);
                \path[fill=white,rounded corners=2] (#2.north west) rectangle (#2.south east);
                \foreach \c/\d in {#1} {
                    \ifthenelse{\equal{\c}{\d}}{
                    }{
                        \path[fill=\d,line width=0] ($(#2\c.north west) + (0,1)$) rectangle ($(#2\c.south east) + (0,-1)$);
                    }
                    \ifthenelse{\equal{\c}{-}}{}{
                        \node[w] at (#2\c){\c};
                    }
                };
                \draw[rounded corners=2] (#2.north west) rectangle (#2.south east);
            \end{scope}
        }
    },
    matchlabel/.style={circle, draw,font=\footnotesize,fill=white,inner sep=.5pt,text=black,minimum height=.8em},
    match/.style={ultra thick,line cap=butt},
    match-substitution/.style={match,draw=substitution},
    match-insertion/.style={match,draw=insertion},
    match-deletion/.style={match,draw=deletion},
    match-correct/.style={match,draw=correct},
}
\newcommand\rowstep{.6em}
\begin{subfigure}[t]{\textwidth}
   
    \centering
    \begin{tikzpicture}
        \coordinate (c0) at (-.5em, 0); %
        \foreach \i in {1,2,...,8} {
            \coordinate (c\i) at (1em*\i, 0);
        }
        \foreach \i in {1, 2, 3} {
            \coordinate (r\i) at (0, -\rowstep*\i);
        }

        \segm{r1}{A/c1,B/c2,C/c3}{ref}{1}
        \segm{r3}{G/c4}{ref}{2}
        \segm{r2}{E/c5,F/c6}{ref}{3}
        \segm{r1}{D/c7}{ref}{4}
        \segm{r3}{H/c8}{ref}{5}

        \begin{pgfonlayer}{bbg}
            \node[transcriptbox,fit={(ref-1)(ref-5)}] (ref) {};
            \draw[gray] 
            (ref.west|-r1) -- (ref.east|-r1)
            (ref.west|-r2) -- (ref.east|-r2)
            (ref.west|-r3) -- (ref.east|-r3);
        \end{pgfonlayer}
        \node[anchor=east] at ($(ref.west) + (-.5em,0)$) {reference};

        \foreach \i in {1, 2, 3} {
            \coordinate (r\i) at (0, -.4em -\rowstep*\i);
        }
        \foreach \i in {1,2,...,8} {
            \coordinate (c\i) at (1em*\i+20em, 0);
        }

        \segm{r2}{A/c1,B/c2}{hyp}{1}
        \segm{r1}{F/c7,H/c8}{hyp}{2}
        \segm{r1}{C/c3,D/c6}{hyp}{3}
        \segm{r2}{E/c5}{hyp}{4}

        \begin{pgfonlayer}{bbg}
            \node[transcriptbox,fit={(hyp-1)(hyp-2)(hyp-3)(hyp-4)}] (hyp) {};
            \draw[gray] 
            (c1|-r1) ++(-1em,0) -- (c8|-r1) -- ++(1em,0)
            (c1|-r2) ++(-1em,0) -- (c8|-r2) -- ++(1em,0);
        \end{pgfonlayer}
        \node[anchor=west] at ($(hyp.east) + (.5em,0)$) {hypothesis};

        \node[draw=black,rounded corners=2,fill=white] at ($(ref.east)!.5!(hyp.west)$) (asr) {Meeting ASR system};
        \draw[arrow] (ref) -- (asr);
        \draw[arrow] (asr) -- (hyp);

    \end{tikzpicture}
    \subcaption{
        Reference transcripts and system outputs. 
        The speech recognizer outputs two hypothesis streams while the ground truth annotations contain three speakers.
        \label{fig:example-sys}
    }
\end{subfigure}
\hspace*{1em}
\begin{subfigure}[t]{.3\textwidth}
    \centering
\begin{tikzpicture}
    \coordinate (c0) at (-.5em, 0); %
    \foreach \i in {1,2,...,10} {
        \coordinate (c\i) at (1em*\i, 0);
    }
    \foreach \i in {0, 1, 2, 3} {
        \coordinate (r\i) at (0, -.6em*\i);
    }

    \segm{r1}{A/c1/correct,B/c2/correct,C/c3/substitution}{ref}{1}
    \segm{r3}{G/c4/deletion}{ref}{2}
    \segm{r2}{E/c5/substitution,F/c6/correct}{ref}{3}
    \segm{r1}{D/c7/deletion}{ref}{4}
    \segm{r3}{H/c8/deletion}{ref}{5}

    \begin{pgfonlayer}{bbg}
        \node[transcriptbox,fit={(ref-1)(ref-5)}] (ref) {};
        \draw[gray] 
        (ref.east|-r1) -- (ref.west|-r1)
        (ref.east|-r2) -- (ref.west|-r2)
        (ref.east|-r3) -- (ref.west|-r3);
    \end{pgfonlayer}
    \node[anchor=east] at ($(ref.west) + (-.5em,0)$) {reference};

    \foreach \i in {1,2,3} {
        \coordinate (ha\i) at ($(r3)+(0,-3em)+(0, -\rowstep*\i)$);
    }
    \segm{ha1}{A/c1/correct,B/c2/correct}{hypa}{1}
    \segm{ha2}{C/c3/insertion,D/c6/substitution}{hypa}{2}
    \segm{ha2}{F/c7/correct,H/c8/insertion}{hypa}{3}
    \segm{ha1}{E/c5/substitution}{hypa}{4}

    \begin{pgfonlayer}{bbg}
        \node[transcriptbox,fit={(hypa-1)(hypa-2)(hypa-3)(hypa-4)($(c1|-ha3) + (0,-.25em)$)}] (hypa) {};
        \draw[gray] 
            (hypa.west|-ha1) -- (hypa.east|-ha1)
            (hypa.west|-ha2) -- (hypa.east|-ha2)
            (hypa.west|-ha3) -- (hypa.east|-ha3);
    \end{pgfonlayer}
    \node[anchor=east] at ($(hypa.west) + (-.5em,0)$) {\tbl{modified\\hypothesis}};

    \foreach \i in {1,2} {
        \coordinate (h\i) at ($(ha3.south)+(0,-3em)+(0, -0\rowstep*\i)$);
    }

    \segm{h2}{A/c1,B/c2}{hyp}{1}
    \segm{h1}{C/c3,D/c6}{hyp}{2}
    \segm{h1}{F/c7,H/c8}{hyp}{3}
    \segm{h2}{E/c5}{hyp}{4}

    \begin{pgfonlayer}{bbg}
        \node[transcriptbox,fit={(hyp-1)(hyp-2)(hyp-3)(hyp-4)}] (hyp) {};
        \draw[gray] 
            (hyp.west|-h1) -- (hyp.east|-h1)
            (hyp.west|-h2) -- (hyp.east|-h2);
    \end{pgfonlayer}
    \node[anchor=east] at ($(hyp.west) + (-.5em,0)$) {hypothesis};

    \coordinate (m) at ($(r3)!.5!(ha1)$);
    \node[matchlabel,fill=correct] (l1) at ($(m-|c1) + (-.75em,0)$) {C};
    \node[matchlabel,fill=correct] (l2) at ($(m-|c2) + (-.75em,0)$) {C};
    \node[matchlabel,fill=insertion] (l3) at ($(m-|c3) + (-.75em,0)$) {I};
    \node[matchlabel,fill=substitution] (l4) at ($(m-|c4) + (-.75em,0)$) {S};
    \node[matchlabel,fill=deletion] (l5) at ($(m-|c5) + (-.75em,0)$) {D};
    \node[matchlabel,fill=substitution] (l6) at ($(m-|c6) + (-.75em,0)$) {S};
    \node[matchlabel,fill=correct] (l7) at ($(m-|c7) + (-.75em,0)$) {C};
    \node[matchlabel,fill=deletion] (l8) at ($(m-|c8) + (-.75em,0)$) {D};
    \node[matchlabel,fill=insertion] (l9) at ($(m-|c9) + (-.75em,0)$) {I};
    \node[matchlabel,fill=deletion] (l10) at ($(m-|c10) + (-.75em,0)$) {D};

    \begin{pgfonlayer}{bg}
    \draw[match-correct,ultra thick] (ref-A.south) -- (l1) -- (hypa-A.north);
    \draw[match-correct,ultra thick] (ref-B.south) -- (l2) -- (hypa-B.north);
    \draw[insertion,ultra thick] (l3) -- (hypa-C.north);
    \draw[match-substitution,ultra thick] (ref-C.south) -- (l4) -- (hypa-E.north);
    \draw[match-deletion,ultra thick] (ref-G.south) -- (l5);
    \draw[match-substitution,ultra thick] (ref-E.south) -- (l6) -- (hypa-D.north);
    \draw[match-correct,ultra thick] (ref-F.south) -- (l7) -- (hypa-F.north);
    \draw[match-deletion,ultra thick] (ref-D.south) -- (l8);
    \draw[insertion,ultra thick] (l9) -- (hypa-H.north);
    \draw[match-deletion,ultra thick] (ref-H.south) -- (l10);
    \node[anchor=east,outer sep=0,inner sep=0] at ($(m-|ref.west) + (-.95em,0)$) (matchinglabel) {Matching};
     \end{pgfonlayer}
    \begin{pgfonlayer}{bbg}
        \node[matchingbox,fit={(l1)(l10)}] (matching) {};
    \end{pgfonlayer}
    
        \node[draw,rounded corners=2] at ($(hyp.north)!.5!(hypa.south)$) (assign-label) {Permute (\cref{eq:cp})};
       \draw[arrow] (hyp) -- (assign-label) -- (hypa);
    \end{tikzpicture}
    
    \subcaption{
        $\mathrm{cpWER}=\frac{\cbox{insertion}{2} + \cbox{deletion}{3} + \cbox{substitution}{2}}{8}=87.5\si{\percent}$.
        The hypothesis streams are reordered with \cref{eq:cp} and empty streams are inserted such that the error rate is minimized.
    }
    \label{fig:example-cp}
\end{subfigure}
\hspace*{1em}
\begin{subfigure}[t]{.3\textwidth}
    \centering
    \begin{tikzpicture}
        \coordinate (c0) at (-.5em, 0); %
        \foreach \i in {1,2,...,8} {
            \coordinate (c\i) at (1em*\i, 0);
        }
        \foreach \i in {0, 1, 2, 3} {
            \coordinate (r\i) at (0, 0*\i);
        }

        \segm{r1}{A/c1,B/c2,C/c3}{ref}{1}
        \segm{r3}{G/c4}{ref}{2}
        \segm{r2}{E/c5,F/c6}{ref}{3}
        \segm{r1}{D/c7}{ref}{4}
        \segm{r3}{H/c8}{ref}{5}

        \begin{pgfonlayer}{bbg}
            \node[transcriptbox,fit={(ref-1)(ref-2)(ref-3)(ref-4)(ref-5)}] (ref) {};
            \draw[gray] 
                (ref.west|-r1) -- (ref.east|-r1)
                (ref.west|-r2) -- (ref.east|-r2)
                (ref.west|-r3) -- (ref.east|-r3);
        \end{pgfonlayer}
        \node[anchor=east] at ($(ref.west) + (-.5em,0)$) {reference};

        \foreach \i in {1,2} {
            \coordinate (ra\i) at ($(r3)+(0,-3.75em)+(0, -\rowstep*\i)$);
        }
        \segm{ra2}{A/c1/correct,B/c2/correct,C/c3/substitution}{refa}{1}
        \segm{ra1}{G/c4/substitution}{refa}{2}
        \segm{ra1}{E/c5/substitution,F/c6/correct}{refa}{3}
        \segm{ra2}{D/c7/deletion}{refa}{4}
        \segm{ra1}{H/c8/correct}{refa}{5}
        \begin{pgfonlayer}{bbg}
            \node[transcriptbox,fit={(refa-1)(refa-5)}] (refa) {};
            \draw[gray] 
                (refa.west|-ra1) -- (refa.east|-ra1)
                (refa.west|-ra2) -- (refa.east|-ra2);
        \end{pgfonlayer}
       \node[anchor=east] at ($(refa.west) + (-.5em,0)$) {\tbl{modified\\reference}};

        \foreach \i in {1,2} {
            \coordinate (h\i) at ($(ra2)+(0,-3em)+(0, -\rowstep*\i)$);
        }

        \segm{h2}{A/c1/correct,B/c2/correct}{hyp}{1}
        \segm{h1}{F/c7/correct,H/c8/correct}{hyp}{2}
        \segm{h1}{C/c3/substitution,D/c6/substitution}{hyp}{3}
        \segm{h2}{E/c5/substitution}{hyp}{4}
        \begin{pgfonlayer}{bbg}
            \node[transcriptbox,fit={(hyp-1)(hyp-2)(hyp-3)(hyp-4)}] (hyp) {};
            \draw[gray] 
                (hyp.east|-h1) -- (hyp.west|-h1)
                (hyp.east|-h2) -- (hyp.west|-h2);
        \end{pgfonlayer}
        \node[anchor=east] at ($(hyp.west) + (-.5em,0)$) {hypothesis};

        \begin{pgfonlayer}{bbg}
        \draw[assign] (ref-1) -- (refa-1);
        \draw[assign] (ref-2) -- (refa-2);
        \draw[assign] (ref-3) -- (refa-3);
        \draw[assign] (ref-4) -- (refa-4);
        \draw[assign] (ref-5) -- (refa-5);
        \end{pgfonlayer}

        \coordinate (m) at ($(ra2)!.5!(h1)$);
        \node[matchlabel,fill=correct] (l1) at (m-|c1) {C};
        \node[matchlabel,fill=correct] (l2) at (m-|c2) {C};
        \node[matchlabel,fill=substitution] (l3) at (m-|c3) {S};
        \node[matchlabel,fill=substitution] (l4) at (m-|c4) {S};
        \node[matchlabel,fill=substitution] (l5) at (m-|c5) {S};
        \node[matchlabel,fill=correct] (l6) at (m-|c6) {C};
        \node[matchlabel,fill=deletion] (l7) at (m-|c7) {D};
        \node[matchlabel,fill=correct] (l8) at (m-|c8) {C};
        \begin{pgfonlayer}{bg}
        \draw[match-correct,ultra thick] (refa-A.south) -- (l1) -- (hyp-A.north);
        \draw[match-correct,ultra thick] (refa-B.south) -- (l2) -- (hyp-B.north);
        \draw[match-substitution,ultra thick] (refa-C.south) -- (l3) -- (hyp-E.north);
        \draw[match-substitution,ultra thick] (refa-G.south) -- (l4) -- (hyp-C.north);
        \draw[match-substitution,ultra thick] (refa-E.south) -- (l5) -- (hyp-D.north);
        \draw[match-correct,ultra thick] (refa-F.south) -- (l6) -- (hyp-F.north);
        \draw[match-deletion,ultra thick] (refa-D.south) -- (l7);
        \draw[match-correct,ultra thick] (refa-H.south) -- (l8) -- (hyp-H.north);
        \node[anchor=east,outer sep=0,inner sep=0] at (m-|c0) (matchinglabel) {Matching};
        \end{pgfonlayer}
        \begin{pgfonlayer}{bbg}
            \node[fit={(l1)(l8)},matchingbox] (matching) {};
        \end{pgfonlayer}
        
        \node[draw,rounded corners=2,fill=white, anchor=east] at ($(refa.north east)!.5!(ref.south east)$) (assign-label) {Modify labels $\rlbl$ (\cref{eq:orc})};

    \end{tikzpicture} 
    \subcaption{
        $\text{ORC-WER}=\frac{\cbox{insertion}{0} + \cbox{deletion}{1} + \cbox{substitution}{3}}{8}=50.0\si{\percent}$.
        The reference labels $\rlbl$ are modified with \cref{eq:orc} to minimize the error rate and preserve the global order of segments.
        \label{fig:example-orc}
    }
\end{subfigure}
\hspace*{1em}
\begin{subfigure}[t]{.3\textwidth}
    \centering
    \begin{tikzpicture}
        \coordinate (c0) at (-.5em, 0); %
        \foreach \i in {1,2,...,8} {
            \coordinate (c\i) at (1em*\i, 0);
        }
        \foreach \i in {0, 1, 2, 3} {
            \coordinate (r\i) at (0, -.6em*\i);
        }
       
        \segm{r1}{A/c1,B/c2,C/c3}{ref}{1}
        \segm{r3}{G/c4}{ref}{2}
        \segm{r2}{E/c5,F/c6}{ref}{3}
        \segm{r1}{D/c7}{ref}{4}
        \segm{r3}{H/c8}{ref}{5}

        \begin{pgfonlayer}{bbg}
            \node[transcriptbox,fit={(ref-1)(ref-5)}] (ref) {};
            \draw[gray] 
                (ref.east|-r1) -- (ref.west|-r1)
                (ref.east|-r2) -- (ref.west|-r2)
                (ref.east|-r3) -- (ref.west|-r3);
        \end{pgfonlayer}
        \node[anchor=east] at ($(ref.west) + (-.5em,0)$) {reference};

        \foreach \i in {1,2} {
            \coordinate (ra\i) at ($(r3)+(0,-3.75em)+(0, -\rowstep*\i)$);
        }
        
        \segm{ra2}{A/c1/correct,B/c2/correct,C/c3/substitution}{refa}{1}
        \segm{ra1}{G/c4/substitution}{refa}{2}
        \segm{ra1}{E/c6/deletion,F/c7/correct}{refa}{3}
        \segm{ra1}{D/c5/correct}{refa}{4}
        \segm{ra1}{H/c8/correct}{refa}{5}
        \begin{pgfonlayer}{bbg}
            \node[transcriptbox,fit={(refa-1)(refa-5)}] (refa) {};
            \draw[gray] 
                (refa.west|-ra1) -- (refa.east|-ra1)
                (refa.west|-ra2) -- (refa.east|-ra2);
        \end{pgfonlayer}
       \node[anchor=east] at ($(refa.west) + (-.5em,0)$) {\tbl{modified\\reference}};

        \foreach \i in {1,2} {
            \coordinate (h\i) at ($(ra2)+(0,-3em)+(0, -\rowstep*\i)$);
        }

        \segm{h2}{A/c1/correct,B/c2/correct}{hyp}{1}
        \segm{h1}{F/c7/correct,H/c8/correct}{hyp}{2}
        \segm{h1}{C/c3/substitution,D/c6/correct}{hyp}{3}
        \segm{h2}{E/c5/substitution}{hyp}{4}
        \begin{pgfonlayer}{bbg}
            \node[transcriptbox,fit={(hyp-1)(hyp-2)(hyp-3)(hyp-4)}] (hyp) {};
            \draw[gray] 
                (hyp.west|-h1) -- (hyp.east|-h1)
                (hyp.west|-h2) -- (hyp.east|-h2);
        \end{pgfonlayer}
        \node[anchor=east] at ($(hyp.west) + (-.5em,0)$) {hypothesis};

        \begin{pgfonlayer}{bbg}
            \draw[assign] (ref-1) -- (refa-1);
            \draw[assign] (ref-2) -- (refa-2);
            \draw[assign] (ref-3) -- (refa-3);
            \draw[assign] (ref-4) -- (refa-4);
            \draw[assign] (ref-5) -- (refa-5);
        \end{pgfonlayer}

        \coordinate (m) at ($(ra2)!.5!(h1)$);
        \node[matchlabel,fill=correct] (l1) at (m-|c1) {C};
        \node[matchlabel,fill=correct] (l2) at (m-|c2) {C};
        \node[matchlabel,fill=substitution] (l3) at (m-|c3) {S};
        \node[matchlabel,fill=substitution] (l4) at (m-|c4) {S};
        \node[matchlabel,fill=correct] (l5) at (m-|c5) {C};
        \node[matchlabel,fill=deletion] (l6) at (m-|c6) {D};
        \node[matchlabel,fill=correct] (l7) at (m-|c7) {C};
        \node[matchlabel,fill=correct] (l8) at (m-|c8) {C};

         \begin{pgfonlayer}{bg}
        \draw[match-correct,ultra thick] (refa-A.south) -- (l1) -- (hyp-A.north);
        \draw[match-correct,ultra thick] (refa-B.south) -- (l2) -- (hyp-B.north);
        \draw[match-substitution,ultra thick] (refa-C.south) -- (l3) -- (hyp-E.north);
        \draw[match-substitution,ultra thick] (refa-G.south) -- (l4) -- (hyp-C.north);
        \draw[match-correct,ultra thick] (refa-D.south) -- (l5) -- (hyp-D.north);
        \draw[match-deletion,ultra thick] (refa-E.south) -- (l6);
        \draw[match-correct,ultra thick] (refa-F.south) -- (l7) -- (hyp-F.north);
        \draw[match-correct,ultra thick] (refa-H.south) -- (l8) -- (hyp-H.north);
         \end{pgfonlayer}
        \node[anchor=east,outer sep=0,inner sep=0] at (m-|c0) (matchinglabel) {Matching};
        \begin{pgfonlayer}{bbg}
            \node[fit={(l1)(l8)},matchingbox] (matching) {};
        \end{pgfonlayer}

        \node[draw,rounded corners=2,fill=white, anchor=east] at ($(refa.north east)!.5!(ref.south east)$) (assign-label) {Modify $\rlbl$ and order (\cref{eq:mimo})};
        \begin{pgfonlayer}{bg}
            \draw[rounded corners=2,fill=white,opacity=.5] 
            (assign-label.north west) rectangle
            ($(assign-label.south east-|c8) + (.5em,0)$);
        \end{pgfonlayer}

    \end{tikzpicture} 
    \subcaption{
        $\text{MIMO-WER}=\frac{\cbox{insertion}{0} + \cbox{deletion}{1} + \cbox{substitution}{2}}{8}=37.5\si{\percent}$.
        The reference labels $\rlbl$ and order are modified to minimize the error rate, but the order within a speaker is preserved.
        \label{fig:example-mimo}
    }
\end{subfigure}
\hspace*{1em}
\begin{subfigure}[t]{.3\textwidth}
    \centering
    \begin{tikzpicture}
        \coordinate (c0) at (-.5em, 0); %
        \foreach \i in {1,2,...,8} {
            \coordinate (c\i) at (1em*\i, 0);
        }
        \foreach \i in {0, 1, 2, 3} {
            \coordinate (r\i) at (0, -.6em*\i);
        }

        \segm{r1}{A/c1/correct,B/c2/correct,C/c3/correct}{ref}{1}
        \segm{r3}{G/c4/substitution}{ref}{2}
        \segm{r2}{E/c5/correct,F/c6/deletion}{ref}{3}
        \segm{r1}{D/c7/correct}{ref}{4}
        \segm{r3}{H/c8/correct}{ref}{5}

        \begin{pgfonlayer}{bbg}
            \node[transcriptbox,fit={(ref-1)(ref-5)}] (ref) {};
            \draw[gray] 
                (ref.west|-r1) -- (ref.east|-r1)
                (ref.west|-r2) -- (ref.east|-r2)
                (ref.west|-r3) -- (ref.east|-r3);
        \end{pgfonlayer}
        \node[anchor=east] at ($(ref.west) + (-.5em,0)$) {reference};

        \foreach \i in {1,2,3} {
            \coordinate (h\i) at ($(r3)+(0,-3em)+(0, -\rowstep*\i)$);
        }
        \segm{h1}{A/c1/correct,B/c2/correct}{hypa}{1}
        \segm{h3}{F/c7/substitution,H/c8/correct}{hypa}{2}
        \segm{h1}{C/c3/correct,D/c6/correct}{hypa}{3}
        \segm{h2}{E/c5/correct}{hypa}{4}
        \begin{pgfonlayer}{bbg}
            \node[transcriptbox,fit={(hypa-1)(hypa-2)(hypa-3)(hypa-4)}] (hypa) {};
            \draw[gray] 
                (hypa.west|-h1) -- (hypa.east|-h1)
                (hypa.west|-h2) -- (hypa.east|-h2)
                (hypa.west|-h3) -- (hypa.east|-h3);
        \end{pgfonlayer}
       \node[anchor=east] at ($(hypa.west) + (-.5em,0)$) {\tbl{modified\\hypothesis}};

        \foreach \i in {1,2} {
            \coordinate (h\i) at ($(h3)+(0,-3.5em)+(0, -\rowstep*\i)$);
        }

        \segm{h2}{A/c1,B/c2}{hyp}{1}
        \segm{h1}{F/c7,H/c8}{hyp}{2}
        \segm{h1}{C/c3,D/c6}{hyp}{3}
        \segm{h2}{E/c5}{hyp}{4}

        \begin{pgfonlayer}{bbg}
            \node[transcriptbox,fit={(hyp-1)(hyp-2)(hyp-3)(hyp-4)}] (hyp) {};
            \draw[gray] 
                (hyp.west|-h1) -- (hyp.east|-h1)
                (hyp.west|-h2) -- (hyp.east|-h2);
        \end{pgfonlayer}
        \node[anchor=east] at ($(hyp.west) + (-.5em,0)$) {hypothesis};

        \draw[assign] (hyp-1) -- (hypa-1);
        \draw[assign] (hyp-2) -- (hypa-2);
        \draw[assign] (hyp-3) -- (hypa-3);
        \draw[assign] (hyp-4) -- (hypa-4);

        \coordinate (m) at ($(ref.south)!.5!(hypa.north)$);
        \node[matchlabel,fill=correct] (l1) at (m-|c1) {C};
        \node[matchlabel,fill=correct] (l2) at (m-|c2) {C};
        \node[matchlabel,fill=correct] (l3) at (m-|c3) {C};
        \node[matchlabel,fill=substitution] (l4) at (m-|c4) {S};
        \node[matchlabel,fill=correct] (l5) at (m-|c5) {C};
        \node[matchlabel,fill=deletion] (l6) at (m-|c6) {D};
        \node[matchlabel,fill=correct] (l7) at (m-|c7) {C};
        \node[matchlabel,fill=correct] (l8) at (m-|c8) {C};

         \begin{pgfonlayer}{bg}
        \draw[match-correct,ultra thick] (ref-A.south) -- (l1) -- (hypa-A.north);
        \draw[match-correct,ultra thick] (ref-B.south) -- (l2) -- (hypa-B.north);
        \draw[match-correct,ultra thick] (ref-C.south) -- (l3) -- (hypa-C.north);
        \draw[match-substitution,ultra thick] (ref-G.south) -- (l4) -- (hypa-F.north);
        \draw[match-correct,ultra thick] (ref-E.south) -- (l5) -- (hypa-E.north);
        \draw[match-deletion,ultra thick] (ref-F.south) -- (l6);
        \draw[match-correct,ultra thick] (ref-D.south) -- (l7) -- (hypa-D.north);
        \draw[match-correct,ultra thick] (ref-H.south) -- (l8) -- (hypa-H.north);
         \end{pgfonlayer}
        \node[anchor=east,outer sep=0,inner sep=0] at (m-|c0) (matchinglabel) {Matching};
        \begin{pgfonlayer}{bbg}
            \node[matchingbox,fit={(l1)(l8)}] (matching) {};
        \end{pgfonlayer}

        \node[draw,rounded corners=2,fill=white, anchor=east] at ($(hyp.north east)!.5!(hypa.south east)$) (assign-label) {Modify labels $\hlbl$ (\cref{eq:di-cp})};
    \end{tikzpicture} 
    \subcaption{
        $\text{DI-cpWER}=\frac{\cbox{insertion}{0} + \cbox{deletion}{1} + \cbox{substitution}{1}}{8}=25.0\si{\percent}$.
        The hypothesis stream labels are modified with \cref{eq:di-cp} to minimize the error rate. The global order of segments is preserved.
        \label{fig:example-di-cp}
    }
\end{subfigure}
\hspace*{1em}
\begin{subfigure}[t]{.3\textwidth}
    \centering
    \begin{tikzpicture}
        \coordinate (c0) at (-.5em, 0); %
        \foreach \i in {1,2,...,8} {
            \coordinate (c\i) at (1em*\i, 0);
        }
        \foreach \i in {0, 1, 2, 3} {
            \coordinate (r\i) at (0, 0*\i);
        }

        \segm{r1}{A/c1}{ref}{1}
        \segm{r1}{B/c2}{ref}{2}
        \segm{r1}{C/c3}{ref}{3}
        \segm{r3}{G/c4}{ref}{4}
        \segm{r2}{E/c5}{ref}{5}
        \segm{r2}{F/c6}{ref}{6}
        \segm{r1}{D/c7}{ref}{7}
        \segm{r3}{H/c8}{ref}{8}

        \begin{pgfonlayer}{bbg}
            \node[transcriptbox,fit={(ref-1)(ref-2)(ref-3)(ref-4)(ref-5)(ref-6)(ref-7)(ref-8)}] (ref) {};
            \draw[gray] 
                (ref.west|-r1) -- (ref.east|-r1)
                (ref.west|-r2) -- (ref.east|-r2)
                (ref.west|-r3) -- (ref.east|-r3);
        \end{pgfonlayer}
        \node[anchor=east] at ($(ref.west) + (-.5em,0)$) {reference};

        \foreach \i in {1,2} {
            \coordinate (ra\i) at ($(r3)+(0,-3.75em)+(0, -\rowstep*\i)$);
        }

        \segm{ra2}{A/c1/correct}{refa}{1}
        \segm{ra2}{B/c2/correct}{refa}{2}
        \segm{ra1}{C/c3/correct}{refa}{3}
        \segm{ra1}{G/c4/substitution}{refa}{4}
        \segm{ra2}{E/c5/correct}{refa}{5}
        \segm{ra1}{F/c6/correct}{refa}{6}
        \segm{ra2}{D/c7/deletion}{refa}{7}
        \segm{ra1}{H/c8/correct}{refa}{8}

        \begin{pgfonlayer}{bbg}
            \node[transcriptbox,fit={(refa-1)(refa-2)(refa-3)(refa-4)(refa-5)(refa-6)(refa-7)(refa-8)}] (refa) {};
            \draw[gray] 
                (refa.west|-ra1) -- (refa.east|-ra1)
                (refa.west|-ra2) -- (refa.east|-ra2);
        \end{pgfonlayer}
       \node[anchor=east] at ($(refa.west) + (-.5em,0)$) {\tbl{modified\\reference}};

        \foreach \i in {1,2} {
            \coordinate (h\i) at ($(ra2)+(0,-3em)+(0, -\rowstep*\i)$);
        }

        \segm{h2}{A/c1/correct}{hyp}{1}
        \segm{h2}{B/c2/correct}{hyp}{2}
        \segm{h1}{F/c7/correct}{hyp}{3}
        \segm{h1}{H/c8/correct}{hyp}{4}
        \segm{h1}{C/c3/correct}{hyp}{5}
        \segm{h1}{D/c6/substitution}{hyp}{6}
        \segm{h2}{E/c5/correct}{hyp}{7}

        \begin{pgfonlayer}{bbg}
            \node[transcriptbox,fit={(hyp-1)(hyp-2)(hyp-3)(hyp-4)(hyp-5)(hyp-6)(hyp-7)}] (hyp) {};
            \draw[gray] 
                (hyp.west|-h1) -- (hyp.east|-h1)
                (hyp.west|-h2) -- (hyp.east|-h2);
        \end{pgfonlayer}
        \node[anchor=east] at ($(hyp.west) + (-.5em,0)$) {hypothesis};

        \begin{pgfonlayer}{bbg}
            \foreach \i in {1,2,3,4,5,6,7,8} {
                \draw[assign] (ref-\i) -- (refa-\i);
            }
        \end{pgfonlayer}

        \coordinate (m) at ($(ra2)!.5!(h1)$);
        \node[matchlabel,fill=correct] (l1) at (m-|c1) {C};
        \node[matchlabel,fill=correct] (l2) at (m-|c2) {C};
        \node[matchlabel,fill=correct] (l3) at (m-|c3) {S};
        \node[matchlabel,fill=substitution] (l4) at (m-|c4) {S};
        \node[matchlabel,fill=correct] (l5) at (m-|c5) {S};
        \node[matchlabel,fill=correct] (l6) at (m-|c6) {C};
        \node[matchlabel,fill=deletion] (l7) at (m-|c7) {D};
        \node[matchlabel,fill=correct] (l8) at (m-|c8) {C};
        
        \begin{pgfonlayer}{bg}
        \draw[match-correct,ultra thick] (refa-A.south) -- (l1.center) -- (hyp-A.north);
        \draw[match-correct,ultra thick] (refa-B.south) -- (l2.center) -- (hyp-B.north);
        \draw[match-correct,ultra thick] (refa-C.south) -- (l3.center) -- (hyp-C.north);
        \draw[match-substitution,ultra thick] (refa-G.south) -- (l4.center) -- (hyp-D.north);
        \draw[match-correct,ultra thick] (refa-E.south) -- (l5.center) -- (hyp-E.north);
        \draw[match-correct,ultra thick] (refa-F.south) -- (l6.center) -- (hyp-F.north);
        \draw[match-deletion,ultra thick] (refa-D.south) -- (l7.center);
        \draw[match-correct,ultra thick] (refa-H.south) -- (l8.center) -- (hyp-H.north);
        \end{pgfonlayer}
        \node[anchor=east,outer sep=0,inner sep=0] at (m-|c0) (matchinglabel) {Matching};
        \begin{pgfonlayer}{bbg}
            \node[fit={(l1)(l8)},matchingbox] (matching) {};
        \end{pgfonlayer}
        
        \node[draw,rounded corners=2,fill=white, anchor=east] at ($(refa.north east)!.5!(ref.south east)$) (assign-label) {Modify labels $\rlbl$ (\cref{eq:orc})};

    \end{tikzpicture}
    \subcaption{
        $\text{wl-ORC-WER}=\frac{\cbox{insertion}{0} + \cbox{deletion}{1} + \cbox{substitution}{1}}{8}=25.0\si{\percent}$.
        The alignment is the same as ORC-WER (\cref{fig:example-orc}), but all segments are broken down into words before matching.
        \label{fig:example-wl-orc}
    }
\end{subfigure}
\hspace*{1em}
\begin{subfigure}[t]{.3\textwidth}
    \centering
    \begin{tikzpicture}
        
        \coordinate (c0) at (-.5em, 0); %
        \foreach \i in {1,2,...,8} {
            \coordinate (c\i) at (1em*\i, 0);
        }
        \foreach \i in {0, 1, 2, 3} {
            \coordinate (r\i) at (0, -.6em*\i);
        }

        \segm{r1}{A/c1/correct}{ref}{1}
        \segm{r1}{B/c2/correct}{ref}{2}
        \segm{r1}{C/c3/correct}{ref}{3}
        \segm{r3}{G/c4/deletion}{ref}{4}
        \segm{r2}{E/c5/correct}{ref}{5}
        \segm{r2}{F/c6/correct}{ref}{6}
        \segm{r1}{D/c7/correct}{ref}{7}
        \segm{r3}{H/c8/correct}{ref}{8}

        \begin{pgfonlayer}{bbg}
            \node[transcriptbox,fit={(ref-1)(ref-2)(ref-3)(ref-4)(ref-5)(ref-6)(ref-7)(ref-8)}] (ref) {};
            \draw[gray] 
                (ref.west|-r1) -- (ref.east|-r1)
                (ref.west|-r2) -- (ref.east|-r2)
                (ref.west|-r3) -- (ref.east|-r3);
        \end{pgfonlayer}
        \node[anchor=east] at ($(ref.west) + (-.5em,0)$) {reference};

        \foreach \i in {1,2,3} {
            \coordinate (h\i) at ($(r3)+(0,-3em)+(0, -\rowstep*\i)$);
        }

        \segm{h1}{A/c1/correct}{hypa}{1}
        \segm{h1}{B/c2/correct}{hypa}{2}
        \segm{h1}{C/c3/correct}{hypa}{3}
        \segm{h2}{E/c5/correct}{hypa}{4}
        \segm{h1}{D/c6/correct}{hypa}{5}
        \segm{h2}{F/c7/correct}{hypa}{6}
        \segm{h3}{H/c8/correct}{hypa}{7}

        \begin{pgfonlayer}{bbg}
            \node[transcriptbox,fit={(hypa-1)(hypa-2)(hypa-3)(hypa-4)(hypa-5)(hypa-6)(hypa-7)}] (hypa) {};
            \draw[gray] 
                (hypa.west|-h1) -- (hypa.east|-h1)
                (hypa.west|-h2) -- (hypa.east|-h2)
                (hypa.west|-h3) -- (hypa.east|-h3);
        \end{pgfonlayer}
       \node[anchor=east] at ($(hypa.west) + (-.5em,0)$) {\tbl{modified\\hypothesis}};

        \foreach \i in {1,2} {
            \coordinate (h\i) at ($(h3)+(0,-3.5em)+(0, -\rowstep*\i)$);
        }

        \segm{h2}{A/c1}{hyp}{1}
        \segm{h2}{B/c2}{hyp}{2}
        \segm{h1}{C/c3}{hyp}{3}
        \segm{h2}{E/c5}{hyp}{4}
        \segm{h1}{D/c6}{hyp}{5}
        \segm{h1}{F/c7}{hyp}{6}
        \segm{h1}{H/c8}{hyp}{7}

        \begin{pgfonlayer}{bbg}
            \node[transcriptbox,fit={(hyp-1)(hyp-2)(hyp-3)(hyp-4)(hyp-5)(hyp-6)(hyp-7)}] (hyp) {};
            \draw[gray] 
                (c1|-h1) ++(-1em,0) -- (c8|-h1) -- ++(1em,0)
                (c1|-h2) ++(-1em,0) -- (c8|-h2) -- ++(1em,0);
        \end{pgfonlayer}
        \node[anchor=east] at ($(hyp.west) + (-.5em,0)$) {hypothesis};

        \foreach \i in {1,2,3,4,5,6,7} {
            \draw[assign] (hyp-\i) -- (hypa-\i);
        }

        \coordinate (m) at ($(ref.south)!.5!(hypa.north)$);
        \node[matchlabel,fill=correct] (l1) at (m-|c1) {C};
        \node[matchlabel,fill=correct] (l2) at (m-|c2) {C};
        \node[matchlabel,fill=correct] (l3) at (m-|c3) {C};
        \node[matchlabel,fill=deletion] (l4) at (m-|c4) {D};
        \node[matchlabel,fill=correct] (l5) at (m-|c5) {C};
        \node[matchlabel,fill=correct] (l6) at (m-|c6) {C};
        \node[matchlabel,fill=correct] (l7) at (m-|c7) {C};
        \node[matchlabel,fill=correct] (l8) at (m-|c8) {C};

        \begin{pgfonlayer}{bg}
        \draw[match-correct,ultra thick] (ref-A.south) -- (l1) -- (hypa-A.north);
        \draw[match-correct,ultra thick] (ref-B.south) -- (l2) -- (hypa-B.north);
        \draw[match-correct,ultra thick] (ref-C.south) -- (l3) -- (hypa-C.north);
        \draw[match-deletion,ultra thick] (ref-G.south) -- (l4);
        \draw[match-correct,ultra thick] (ref-E.south) -- (l5) -- (hypa-E.north);
        \draw[match-correct,ultra thick] (ref-F.south) -- (l6) -- (hypa-F.north);
        \draw[match-correct,ultra thick] (ref-D.south) -- (l7) -- (hypa-D.north);
        \draw[match-correct,ultra thick] (ref-H.south) -- (l8) -- (hypa-H.north);
        \end{pgfonlayer}
        \begin{pgfonlayer}{bbg}
            \node[matchingbox,fit={(l1)(l8)}] (matching) {};
        \end{pgfonlayer}
        \node[anchor=east,outer sep=0,inner sep=0] at (m-|c0) (matchinglabel) {Matching};

        \node[draw,rounded corners=2,fill=white, anchor=east] at ($(hyp.north east)!.5!(hypa.south east)$) (assign-label) {Modify labels $\hlbl$ (\cref{eq:di-cp})};
    
    \end{tikzpicture}
    \subcaption{$\text{wl-DI-cpWER}=\frac{\cbox{insertion}{0} + \cbox{deletion}{1} + \cbox{substitution}{0}}{8}=12.5\si{\percent}$.
    The alignment is the same as DI-cpWER (\cref{fig:example-di-cp}), but all segments are broken down into words before matching.
    \label{fig:example-wl-di-cp}
    }
\end{subfigure}

%% file: plots/ex3.tex
\begin{tikzpicture}

\newcommand{\ctext}[1]{\contour{white}{#1}}
\newcommand{\fgtext}[1]{
    \begin{pgfonlayer}{fg}
        \node[viztext] at (tmp) {\ctext{#1}};
    \end{pgfonlayer}
}

\begin{axis}[
    ylabel={Time (s)},
    xmin=-.05, xmax=1.05,
    ymin=634, ymax=639.3,
    ymajorgrids=true,
    grid style=dashed,
    y dir=reverse,
    xtick={.2,.8},
    xticklabels={Reference,Hypothesis},
    height=9cm,
    width=\columnwidth,
    extra description/.code={
        \node[anchor=south] at (.5,1) {\small cpWER: \SI{64.5}{\percent}};
    }
]
\draw[fill=substitution] (0.6, 634.1240000000000000000000000) rectangle (1.0, 635.0162242647058823529411765) coordinate[pos=.5] (tmp); \fgtext{portugese}
\draw[match-substitution] (0.4, 604.9653333333333333333333335) -- (0.6, 634.570112132352941176470588);
\draw[fill=substitution] (0.6, 635.0162242647058823529411765) rectangle (1.0, 635.6110404411764705882352941) coordinate[pos=.5] (tmp); \fgtext{family}
\draw[match-substitution] (0.4, 605.1793333333333333333333335) -- (0.6, 635.3136323529411764705882355);
\draw[fill=substitution] (0.6, 635.6110404411764705882352941) rectangle (1.0, 636.2058566176470588235294118) coordinate[pos=.5] (tmp); \fgtext{though}
\draw[match-substitution] (0.4, 605.322000000000000000000000) -- (0.6, 635.908448529411764705882353);
\draw[fill=substitution] (0.6, 636.2058566176470588235294118) rectangle (1.0, 636.3049926470588235294117647) coordinate[pos=.5] (tmp); \fgtext{i}
\draw[match-substitution] (0.4, 605.4884444444444444444444445) -- (0.6, 636.255424632352941176470588);
\draw[fill=substitution] (0.6, 636.3049926470588235294117647) rectangle (1.0, 636.7015367647058823529411765) coordinate[pos=.5] (tmp); \fgtext{feel}
\draw[match-substitution] (0.4, 605.726222222222222222222222) -- (0.6, 636.5032647058823529411764705);
\draw[fill=substitution] (0.6, 636.7015367647058823529411765) rectangle (1.0, 637.0980808823529411764705882) coordinate[pos=.5] (tmp); \fgtext{like}
\draw[match-substitution] (0.4, 605.8926666666666666666666665) -- (0.6, 636.8998088235294117647058825);
\draw[fill=correct] (0.6, 637.0980808823529411764705882) rectangle (1.0, 637.6928970588235294117647059) coordinate[pos=.5] (tmp); \fgtext{that's}
\draw[match-correct] (0.4, 606.9864444444444444444444445) -- (0.6, 637.395488970588235294117647);
\draw[fill=substitution] (0.6, 637.6928970588235294117647059) rectangle (1.0, 638.0894411764705882352941176) coordinate[pos=.5] (tmp); \fgtext{part}
\draw[match-substitution] (0.4, 607.224222222222222222222222) -- (0.6, 637.891169117647058823529412);
\draw[fill=correct] (0.6, 638.0894411764705882352941176) rectangle (1.0, 638.2877132352941176470588235) coordinate[pos=.5] (tmp); \fgtext{of}
\draw[match-correct] (0.4, 607.366888888888888888888889) -- (0.6, 638.1885772058823529411764705);
\draw[fill=substitution] (0.6, 638.2877132352941176470588235) rectangle (1.0, 638.4859852941176470588235294) coordinate[pos=.5] (tmp); \fgtext{it}
\draw[match-substitution] (0.4, 607.652222222222222222222222) -- (0.6, 638.3868492647058823529411765);
\draw[fill=correct] (0.6, 638.4859852941176470588235294) rectangle (1.0, 638.7833933823529411764705882) coordinate[pos=.5] (tmp); \fgtext{you}
\draw[match-correct] (0.4, 629.972) -- (0.6, 638.634689338235294117647059);
\draw[fill=correct] (0.6, 638.7833933823529411764705882) rectangle (1.0, 639.1799375000000000000000000) coordinate[pos=.5] (tmp); \fgtext{know}
\draw[match-correct] (0.4, 630.140) -- (0.6, 638.981665441176470588235294);
\draw[fill=substitution] (0, 634.4600000000000000000000000) rectangle (0.4, 635.0333333333333333333333333) coordinate[pos=.5] (tmp); \fgtext{portuguese}
\draw[fill=substitution] (0, 635.0333333333333333333333333) rectangle (0.4, 635.3773333333333333333333333) coordinate[pos=.5] (tmp); \fgtext{family}
\draw[fill=substitution] (0, 635.3773333333333333333333333) rectangle (0.4, 635.7213333333333333333333333) coordinate[pos=.5] (tmp); \fgtext{though}
\draw[fill=correct] (0, 635.7213333333333333333333333) rectangle (0.4, 635.7786666666666666666666667) coordinate[pos=.5] (tmp); \fgtext{i}
\draw[match-correct] (0.4, 635.750000000000000000000000) -- (0.6, 645.9720763449367088607594935);
\draw[fill=substitution] (0, 635.7786666666666666666666667) rectangle (0.4, 636.0080000000000000000000000) coordinate[pos=.5] (tmp); \fgtext{feel}
\draw[match-substitution] (0.4, 635.8933333333333333333333335) -- (0.6, 646.3050359968354430379746835);
\draw[fill=substitution] (0, 636.0080000000000000000000000) rectangle (0.4, 636.2373333333333333333333333) coordinate[pos=.5] (tmp); \fgtext{like}
\draw[match-substitution] (0.4, 636.1226666666666666666666665) -- (0.6, 646.8044754746835443037974685);
\draw[fill=substitution] (0, 636.2373333333333333333333333) rectangle (0.4, 636.5813333333333333333333333) coordinate[pos=.5] (tmp); \fgtext{that's}
\draw[match-substitution] (0.4, 636.4093333333333333333333335) -- (0.6, 647.470394778481012658227848);
\draw[fill=substitution] (0, 636.5813333333333333333333333) rectangle (0.4, 636.8106666666666666666666667) coordinate[pos=.5] (tmp); \fgtext{part}
\draw[match-substitution] (0.4, 636.696000000000000000000000) -- (0.6, 648.0253275316455696202531645);
\draw[fill=substitution] (0, 636.8106666666666666666666667) rectangle (0.4, 636.9253333333333333333333333) coordinate[pos=.5] (tmp); \fgtext{of}
\draw[match-substitution] (0.4, 636.868000000000000000000000) -- (0.6, 648.3582871835443037974683545);
\draw[fill=substitution] (0, 636.9253333333333333333333333) rectangle (0.4, 637.0400000000000000000000000) coordinate[pos=.5] (tmp); \fgtext{it}
\draw[match-substitution] (0.4, 636.9826666666666666666666665) -- (0.6, 648.6912468354430379746835445);
\draw[match-substitution] (0.4, 630.332) -- (0.6, 642.6979731012658227848101265);
\draw[match-substitution] (0.4, 630.596) -- (0.6, 643.141919303797468354430380);
\draw[match-substitution] (0.4, 630.932) -- (0.6, 643.530372231012658227848101);
\draw[match-substitution] (0.4, 634.7466666666666666666666665) -- (0.6, 643.8078386075949367088607595);
\draw[match-substitution] (0.4, 635.2053333333333333333333335) -- (0.6, 644.029811708860759493670886);
\draw[match-substitution] (0.4, 635.5493333333333333333333335) -- (0.6, 644.362771360759493670886076);
\draw[match-substitution] (0.4, 643.220) -- (0.6, 648.913219936708860759493671);
\end{axis}
\end{tikzpicture}

%% file: plots/ex1.tex
\newcommand{\ctext}[1]{\contour{white}{#1}}
\newcommand{\fgtext}[1]{
    \begin{pgfonlayer}{fg}
        \node[viztext] at (tmp) {\ctext{#1}};
    \end{pgfonlayer}
}

\begin{tikzpicture}

    \begin{axis}[
        ylabel={Time (s)},
        xmin=-.05, xmax=1.05,
        ymin=634, ymax=639.3,
        ymajorgrids=true,
        grid style=dashed,
        y dir=reverse,
        xtick={.2,.8},
        xticklabels={Reference,Hypothesis},
        height=9cm,
        width=\columnwidth,
        extra description/.code={
            \node[anchor=south] at (.5,1) {\small tcpWER: \SI{70.9}{\percent}};
        }
    ]
\draw[fill=substitution] (0.6, 634.1240000000000000000000000) rectangle (1.0, 635.0162242647058823529411765) coordinate[pos=.5] (tmp); \fgtext{portugese}
\draw[match-substitution] (0.4, 629.972) -- (0.6, 634.570112132352941176470588);
\draw[fill=correct] (0.6, 635.0162242647058823529411765) rectangle (1.0, 635.6110404411764705882352941) coordinate[pos=.5] (tmp); \fgtext{family}
\draw[fill=correct] (0.6, 635.6110404411764705882352941) rectangle (1.0, 636.2058566176470588235294118) coordinate[pos=.5] (tmp); \fgtext{though}
\draw[match-correct] (0.4, 635.5493333333333333333333335) -- (0.6, 635.908448529411764705882353);
\draw[fill=correct] (0.6, 636.2058566176470588235294118) rectangle (1.0, 636.3049926470588235294117647) coordinate[pos=.5] (tmp); \fgtext{i}
\draw[match-correct] (0.4, 635.750000000000000000000000) -- (0.6, 636.255424632352941176470588);
\draw[fill=correct] (0.6, 636.3049926470588235294117647) rectangle (1.0, 636.7015367647058823529411765) coordinate[pos=.5] (tmp); \fgtext{feel}
\draw[match-correct] (0.4, 635.8933333333333333333333335) -- (0.6, 636.5032647058823529411764705);
\draw[fill=correct] (0.6, 636.7015367647058823529411765) rectangle (1.0, 637.0980808823529411764705882) coordinate[pos=.5] (tmp); \fgtext{like}
\draw[match-correct] (0.4, 636.1226666666666666666666665) -- (0.6, 636.8998088235294117647058825);
\draw[fill=correct] (0.6, 637.0980808823529411764705882) rectangle (1.0, 637.6928970588235294117647059) coordinate[pos=.5] (tmp); \fgtext{that's}
\draw[match-correct] (0.4, 636.4093333333333333333333335) -- (0.6, 637.395488970588235294117647);
\draw[fill=correct] (0.6, 637.6928970588235294117647059) rectangle (1.0, 638.0894411764705882352941176) coordinate[pos=.5] (tmp); \fgtext{part}
\draw[match-correct] (0.4, 636.696000000000000000000000) -- (0.6, 637.891169117647058823529412);
\draw[fill=correct] (0.6, 638.0894411764705882352941176) rectangle (1.0, 638.2877132352941176470588235) coordinate[pos=.5] (tmp); \fgtext{of}
\draw[match-correct] (0.4, 636.868000000000000000000000) -- (0.6, 638.1885772058823529411764705);
\draw[fill=correct] (0.6, 638.2877132352941176470588235) rectangle (1.0, 638.4859852941176470588235294) coordinate[pos=.5] (tmp); \fgtext{it}
\draw[match-correct] (0.4, 636.9826666666666666666666665) -- (0.6, 638.3868492647058823529411765);
\draw[fill=insertion] (0.6, 638.4859852941176470588235294) rectangle (1.0, 638.7833933823529411764705882) coordinate[pos=.5] (tmp); \fgtext{you}
\draw[fill=insertion] (0.6, 638.7833933823529411764705882) rectangle (1.0, 639.1799375000000000000000000) coordinate[pos=.5] (tmp); \fgtext{know}
\draw[fill=deletion] (0, 634.4600000000000000000000000) rectangle (0.4, 635.0333333333333333333333333) coordinate[pos=.5] (tmp); \fgtext{portuguese}
\draw[fill=correct] (0, 635.0333333333333333333333333) rectangle (0.4, 635.3773333333333333333333333) coordinate[pos=.5] (tmp); \fgtext{family}
\draw[match-correct] (0.4, 635.2053333333333333333333335) -- (0.6, 635.3136323529411764705882355);
\draw[fill=correct] (0, 635.3773333333333333333333333) rectangle (0.4, 635.7213333333333333333333333) coordinate[pos=.5] (tmp); \fgtext{though}
\draw[fill=correct] (0, 635.7213333333333333333333333) rectangle (0.4, 635.7786666666666666666666667) coordinate[pos=.5] (tmp); \fgtext{i}
\draw[fill=correct] (0, 635.7786666666666666666666667) rectangle (0.4, 636.0080000000000000000000000) coordinate[pos=.5] (tmp); \fgtext{feel}
\draw[fill=correct] (0, 636.0080000000000000000000000) rectangle (0.4, 636.2373333333333333333333333) coordinate[pos=.5] (tmp); \fgtext{like}
\draw[fill=correct] (0, 636.2373333333333333333333333) rectangle (0.4, 636.5813333333333333333333333) coordinate[pos=.5] (tmp); \fgtext{that's}
\draw[fill=correct] (0, 636.5813333333333333333333333) rectangle (0.4, 636.8106666666666666666666667) coordinate[pos=.5] (tmp); \fgtext{part}
\draw[fill=correct] (0, 636.8106666666666666666666667) rectangle (0.4, 636.9253333333333333333333333) coordinate[pos=.5] (tmp); \fgtext{of}
\draw[fill=correct] (0, 636.9253333333333333333333333) rectangle (0.4, 637.0400000000000000000000000) coordinate[pos=.5] (tmp); \fgtext{it} 
    \end{axis}
    \end{tikzpicture}

%% file: plots/ex2.tex
\pgfkeys{/pgf/number format/.cd,1000 sep={\,}}  %
\newcommand{\ctext}[1]{\contour{white}{#1}}
\newcommand{\fgtext}[1]{
    \begin{pgfonlayer}{fg}
        \node[viztext] at (tmp) {\ctext{#1}};
    \end{pgfonlayer}
}
\begin{tikzpicture}
\begin{axis}[
    ylabel={Time (s)},
    xmin=-.05, xmax=1.05,
    ymin=1949.9, ymax=1959,
    ymajorgrids=true,
    grid style=dashed,
    y dir=reverse,
    xtick={.2,.8},
    xticklabels={Reference,Hypothesis},
    height=9cm,
    width=\columnwidth,
    extra description/.code={
        \node[anchor=south] at (.45,1) {\small cpWER: \SI{38.3}{\percent}, tcpWER: \SI{51.0}{\percent}};
    }
]
\draw[fill=correct] (0, 1949.780285714285714285714286) rectangle (0.4, 1950.195142857142857142857143) coordinate[pos=.5] (tmp); \fgtext{very}
\draw[match-correct] (0.4, 1949.987714285714285714285714) -- (0.6, 1947.646470588235294117647059);
\draw[fill=correct] (0, 1950.195142857142857142857143) rectangle (0.4, 1950.610000000000000000000000) coordinate[pos=.5] (tmp); \fgtext{cool}
\draw[match-correct] (0.4, 1950.402571428571428571428572) -- (0.6, 1948.015882352941176470588236);
\draw[fill=correct] (0.6, 1949.880) rectangle (1.0, 1950.270) coordinate[pos=.5] (tmp); \fgtext{harder}
\draw[fill=correct] (0.6, 1950.270) rectangle (1.0, 1950.530) coordinate[pos=.5] (tmp); \fgtext{here}
\draw[fill=correct] (0.6, 1953.460000000000000000000000) rectangle (1.0, 1953.680000000000000000000000) coordinate[pos=.5] (tmp); \fgtext{though}
\draw[fill=correct] (0.6, 1953.680000000000000000000000) rectangle (1.0, 1953.753333333333333333333333) coordinate[pos=.5] (tmp); \fgtext{on}
\draw[fill=correct] (0.6, 1953.753333333333333333333333) rectangle (1.0, 1953.863333333333333333333333) coordinate[pos=.5] (tmp); \fgtext{the}
\draw[fill=correct] (0.6, 1953.863333333333333333333333) rectangle (1.0, 1954.010000000000000000000000) coordinate[pos=.5] (tmp); \fgtext{west}
\draw[fill=correct] (0.6, 1954.010000000000000000000000) rectangle (1.0, 1954.193333333333333333333333) coordinate[pos=.5] (tmp); \fgtext{coast}
\draw[fill=correct] (0.6, 1954.193333333333333333333333) rectangle (1.0, 1954.340000000000000000000000) coordinate[pos=.5] (tmp); \fgtext{yeah}
\draw[fill=correct] (0, 1953.550) rectangle (0.4, 1954.350) coordinate[pos=.5] (tmp); \fgtext{yeah}
\draw[match-correct] (0.4, 1953.950) -- (0.6, 1948.385294117647058823529412);
\draw[fill=correct] (0, 1955.210000000000000000000000) rectangle (0.4, 1955.451176470588235294117647) coordinate[pos=.5] (tmp); \fgtext{it's}
\draw[match-correct] (0.4, 1955.330588235294117647058824) -- (0.6, 1949.750);
\draw[fill=correct] (0, 1955.451176470588235294117647) rectangle (0.4, 1955.812941176470588235294118) coordinate[pos=.5] (tmp); \fgtext{harder}
\draw[match-correct] (0.4, 1955.632058823529411764705882) -- (0.6, 1950.075);
\draw[fill=correct] (0, 1955.812941176470588235294118) rectangle (0.4, 1956.054117647058823529411765) coordinate[pos=.5] (tmp); \fgtext{here}
\draw[match-correct] (0.4, 1955.933529411764705882352942) -- (0.6, 1950.400);
\draw[fill=correct] (0, 1956.054117647058823529411765) rectangle (0.4, 1956.415882352941176470588235) coordinate[pos=.5] (tmp); \fgtext{though}
\draw[match-correct] (0.4, 1956.235000000000000000000000) -- (0.6, 1953.570000000000000000000000);
\draw[fill=correct] (0, 1956.415882352941176470588235) rectangle (0.4, 1956.536470588235294117647059) coordinate[pos=.5] (tmp); \fgtext{on}
\draw[match-correct] (0.4, 1956.476176470588235294117647) -- (0.6, 1953.716666666666666666666666);
\draw[fill=correct] (0, 1956.536470588235294117647059) rectangle (0.4, 1956.717352941176470588235294) coordinate[pos=.5] (tmp); \fgtext{the}
\draw[match-correct] (0.4, 1956.626911764705882352941176) -- (0.6, 1953.808333333333333333333333);
\draw[fill=correct] (0, 1956.717352941176470588235294) rectangle (0.4, 1956.958529411764705882352941) coordinate[pos=.5] (tmp); \fgtext{west}
\draw[match-correct] (0.4, 1956.837941176470588235294118) -- (0.6, 1953.936666666666666666666666);
\draw[fill=correct] (0, 1956.958529411764705882352941) rectangle (0.4, 1957.260000000000000000000000) coordinate[pos=.5] (tmp); \fgtext{coast}
\draw[match-correct] (0.4, 1957.109264705882352941176470) -- (0.6, 1954.101666666666666666666666);
\draw[fill=correct] (0, 1958.450) rectangle (0.4, 1959.210) coordinate[pos=.5] (tmp); \fgtext{yeah}
\draw[match-correct] (0.4, 1958.830) -- (0.6, 1954.266666666666666666666666);
\draw[] (0.6, 1955.24) rectangle (1.0, 1957.3) node[viztext-empty] {(empty)};
\draw[] (0.6, 1958.46) rectangle (1.0, 1959.1) node[viztext-empty] {(empty)};
\end{axis}
\end{tikzpicture}

%% file: plots/cp-vs-tcp.tex
\begin{tikzpicture}
    \begin{axis}[
        xlabel={cpWER (\%)},
        ylabel={\tbl{$\Delta$WER (\%)\\cpWER$ - $tcpWER}},
        xmin=0, xmax=80,
        ymin=0, ymax=15,
        grid=major,
        height=5cm,
        width=\columnwidth - 1.25em,
    ]

    \addplot[
        only marks, %
        mark=square,
        color=palette-1,
        draw opacity=0.5,
        ]
        table [x=cpwer,y=difftcpcp,col sep=comma] {data/metrics.csv};

    \node[fill=red,inner sep=1mm,label={[fill=white,inner sep=0,font=\footnotesize]\cref{fig:cp-tcp-ex3,fig:cp-tcp-ex1}}] at (64.5184293800419, 6.43134108838079) {};
    \node[fill=red,inner sep=1mm,label={[fill=white,inner sep=0,font=\footnotesize]\cref{fig:cp-tcp-ex2}}] at (38.1829268292683, 12.829268292682926) {};

    \end{axis}
\end{tikzpicture}

%% file: plots/orc-vs-mimo.tex
\begin{tikzpicture}

    \begin{axis}[
        xlabel={tcMIMO-WER (\%)},
        ylabel={\tbl{$\Delta$WER (\%) \\ $\text{tcORC} - \text{tcMIMO}$}},
        xmin=0, xmax=80,
        ymin=0, ymax=.9,
        grid=major,
        height=5cm,
        width=\columnwidth-1.25em,
    ]

    \addplot[
        only marks, %
        mark=square,
        color=palette-1,
        draw opacity=0.5,
        ]
        table [x=tcmimower,y=difftcorctcmimo,col sep=comma] {data/metrics.csv};
    \end{axis}
\end{tikzpicture}

%% file: plots/timestamp-precision.tex
\begin{tikzpicture}
    \begin{axis}[
        xlabel={$|\Delta_t|$ (\si{\second})},
        ylabel={WER (\%)},
        legend pos=north west,
        xmin=1, xmax=15,
        ymin=0, ymax=100,
        legend style={
            nodes={scale=0.75, transform shape},
            at={(0.99,0.125)},
            anchor=south east
        },
        height=5cm,
        width=\columnwidth,
        grid=major,
    ]

    \addplot[
        mark=*,
        color=palette-3,
        draw opacity=0.5,
        ]
        table [x=amount,y=der,col sep=comma] {data/timestamp_precision.csv};
        \addlegendentry{DER}
        
    \addplot[
        mark=square,
        color=palette-2,
        draw opacity=0.5,
        ]
        table [x=amount,y=tcpwer,col sep=comma] {data/timestamp_precision.csv};
        \addlegendentry{tcpWER}

    \addplot[
            mark=o,
            color=palette-1,
            draw opacity=0.5,
            ]
            table [x=amount,y=cpwer,col sep=comma] {data/timestamp_precision.csv};
            \addlegendentry{cpWER}

    \addplot[
            mark=+,
            color=palette-5,
            draw opacity=0.5,
            ]
            table [x=amount,y=tcorcwer,col sep=comma] {data/timestamp_precision.csv};
            \addlegendentry{tcORC-WER}

    \addplot[
            mark=diamond,
            color=palette-4,
            draw opacity=0.5,
            ]
            table [x=amount,y=di-tcpwer,col sep=comma] {data/timestamp_precision.csv};
            \addlegendentry{DI-tcpWER}

        \draw[black!50, dashed, thin] (5,0) --node[sloped,fill=white,pos=0.9, inner sep=2pt]{$c_\text{tcp}$} (5,100);
    \end{axis}
\end{tikzpicture}

%% file: plots/label-switches.tex
\begin{tikzpicture}
    \begin{axis}[
        xlabel={amount of wrong labels (\si{\percent})},
        ylabel={WER (\%)},
        legend pos=north west,
        xmin=0, xmax=1,
        ymin=0, ymax=100,
        legend style={nodes={scale=0.75, transform shape}},
        height=5cm,
        width=\columnwidth,
        grid=major,
        xtick={0, 0.2, 0.4, 0.6, 0.8, 1},
        xticklabels={0, 20, 40, 60, 80, 100}
    ]

        \addplot[
            mark=*,
            color=palette-3,
            draw opacity=0.5,
            ]
            table [x=amount,y=der,col sep=comma] {data/label_switches.csv};
            \addlegendentry{DER}
        
    \addplot[
        mark=square,
        color=palette-2,
        draw opacity=0.5,
        ]
        table [x=amount,y=tcpwer,col sep=comma] {data/label_switches.csv};
        \addlegendentry{tcpWER}

    \addplot[
            mark=o,
            color=palette-1,
            draw opacity=0.5,
            ]
            table [x=amount,y=cpwer,col sep=comma] {data/label_switches.csv};
            \addlegendentry{cpWER}
        
        \draw[black!50, dashed, thin] (5,0) --node[sloped,fill=white,pos=0.1]{$c_\text{tcp}$} (5,100);

    \addplot[
        mark=+,
        color=palette-5,
        draw opacity=0.5,
        ]
        table [x=amount,y=tcorcwer,col sep=comma] {data/label_switches.csv};
        \addlegendentry{tcORC-WER}
    
    \addplot[
        mark=diamond,
        color=palette-4,
        draw opacity=0.5,
        ]
        table [x=amount,y=di-tcpwer,col sep=comma] {data/label_switches.csv};
        \addlegendentry{DI-tcpWER}
    \end{axis}
\end{tikzpicture}%

%% file: plots/ditcp-vs-invtcorc.tex
\begin{tikzpicture}
    \begin{axis}[
        xlabel={exact DI-tcpWER (\%)},
        ylabel={\tbl{$\Delta$ORC-WER (\%)\\greedy$ - $exact}},
        xmin=10, xmax=80,
        ymin=0, ymax=.4,
        grid=major,
        height=5cm,
        width=\columnwidth-1.25em,
    ]

    \addplot[
        only marks, %
        mark=square,
        color=palette-1,
        draw opacity=0.5,
        ]
        table [x=ditcpwer,y=diffditcpinvtcorc,col sep=comma] {data/metrics.csv};
    \end{axis}
\end{tikzpicture}

%% file: plots/benchmark.tex
\begin{tikzpicture}

\definecolor{darkseagreen117191113}{RGB}{117,191,113}
\definecolor{darkslategray38}{RGB}{38,38,38}
\definecolor{dimgray84}{RGB}{84,84,84}
\definecolor{indianred199109109}{RGB}{199,109,109}
\definecolor{lightgray204}{RGB}{204,204,204}
\definecolor{lightslategray147116171}{RGB}{147,116,171}
\definecolor{peru21713894}{RGB}{217,138,94}
\definecolor{steelblue89124191}{RGB}{89,124,191}

\begin{axis}[
legend cell align={left},
legend style={
  at={(1,0)},
  anchor=south east,
},
xtick={0,1,2},
xticklabels={\tab{segments\\(2 min)},\tab{raw\\(10 min)},\tab{1h\\(60 min)}},
log basis y={10},
xlabel=LibriCSS segmentation,
xmin=-0.54, xmax=2.54,
ylabel=Execution time per example (s),
ymajorgrids,
ymin=0.001, ymax=1000,
yminorgrids,
ymode=log,
        height=6.5cm,
        width=\columnwidth,
        legend style={nodes={scale=0.75, transform shape}},
]
\draw[draw=white,fill=steelblue89124191] (axis cs:0,0) rectangle (axis cs:0,0);
\addlegendimage{ybar,ybar legend,draw=white,fill=steelblue89124191}
\addlegendentry{MIMO}

\draw[draw=white,fill=peru21713894] (axis cs:0,0) rectangle (axis cs:0,0);
\addlegendimage{ybar,ybar legend,draw=white,fill=peru21713894}
\addlegendentry{tcMIMO}

\draw[draw=white,fill=darkseagreen117191113] (axis cs:0,0) rectangle (axis cs:0,0);
\addlegendimage{ybar,ybar legend,draw=white,fill=darkseagreen117191113}
\addlegendentry{ORC}

\draw[draw=white,fill=indianred199109109] (axis cs:0,0) rectangle (axis cs:0,0);
\addlegendimage{ybar,ybar legend,draw=white,fill=indianred199109109}
\addlegendentry{greedy ORC}

\draw[draw=white,fill=lightslategray147116171] (axis cs:0,0) rectangle (axis cs:0,0);
\addlegendimage{ybar,ybar legend,draw=white,fill=lightslategray147116171}
\addlegendentry{tcORC}

\path [draw=dimgray84, fill=steelblue89124191]
(axis cs:-0.4,0.0313465141225606)
--(axis cs:-0.24,0.0313465141225606)
--(axis cs:-0.24,0.542435014946386)
--(axis cs:-0.4,0.542435014946386)
--(axis cs:-0.4,0.0313465141225606)
--cycle;
\addplot [dimgray84, forget plot]
table {%
-0.32 0.0313465141225606
-0.32 0.0013506359420716
};
\addplot [dimgray84, forget plot]
table {%
-0.32 0.542435014946386
-0.32 1.30363886645064
};
\addplot [dimgray84, forget plot]
table {%
-0.36 0.0013506359420716
-0.28 0.0013506359420716
};
\addplot [dimgray84, forget plot]
table {%
-0.36 1.30363886645064
-0.28 1.30363886645064
};
\addplot [black, mark=o, mark size=2, mark options={solid,fill opacity=0,draw=dimgray84}, only marks, forget plot]
table {%
-0.32 4.28283649785444
-0.32 7.86972165154293
-0.32 2.63555353712291
-0.32 1.46812835149467
-0.32 42.8001189682633
-0.32 1.44628924215212
-0.32 10.721269018203
-0.32 1.35225505298004
-0.32 1.62572459895164
-0.32 7.6848875304684
-0.32 3.58476729961112
-0.32 2.52949200328439
-0.32 1.58409505626187
-0.32 1.64732592310756
-0.32 148.024264151603
-0.32 2.05086010321975
-0.32 2.40405372902751
-0.32 2.15036782026291
-0.32 1.96158030992374
-0.32 1.56572332056239
-0.32 5.28475109580904
-0.32 2.23890073345974
-0.32 2.21533229649067
-0.32 3.92247546818107
-0.32 2.21719832587987
-0.32 101.808884100243
-0.32 925.657853873167
-0.32 1.62679489012808
-0.32 3.31612352961674
-0.32 1.86365842213854
-0.32 1.31575996037573
-0.32 23.4645460621454
-0.32 4.53789694746956
-0.32 1.44522786652669
-0.32 2.06709845550358
-0.32 2.26936391452327
-0.32 75.4729030475952
-0.32 2.2897079157643
-0.32 29.5701670064591
-0.32 1.8725884350948
-0.32 1.58361743325368
-0.32 3.05986627768725
-0.32 2.58462558565661
-0.32 2.1569344939664
-0.32 2.02217501392588
-0.32 1.53974324129522
-0.32 3.36260120328516
-0.32 16.7818348877132
-0.32 516.781313915551
-0.32 2.37601503236219
-0.32 1.45965845929459
-0.32 7.98242643345147
-0.32 3.88252697130665
-0.32 23.1448028172366
-0.32 3.31253806902096
-0.32 3.60873441817239
-0.32 208.36205098033
-0.32 14.922659267392
-0.32 1.7709490634501
-0.32 3.06016010427847
-0.32 5.70270051611587
-0.32 6.29099177485332
-0.32 1.75022465614602
-0.32 6.37175726583227
-0.32 3.28631413830444
-0.32 1.7356350177899
-0.32 1.32317438973114
-0.32 6.10998417520896
-0.32 199.500898000319
-0.32 18.7561039895751
-0.32 2.82599907694385
-0.32 21.9575145230629
-0.32 2.61709500579163
-0.32 6.17833010498434
-0.32 1.85937276026234
-0.32 3.52166783679277
-0.32 2.11055152565241
-0.32 1.97855623653158
-0.32 2.08006210234016
-0.32 1.36207601893693
-0.32 2.10530509203672
-0.32 5.45402095066383
-0.32 1.75534385954961
-0.32 1.76643907194957
-0.32 1.82276188554242
-0.32 1.37730303518474
-0.32 2.81231846073642
-0.32 2.32091902196407
-0.32 2.46426285132766
-0.32 11.2554887820967
-0.32 2.77169662434608
-0.32 3.29621935654432
-0.32 1.46565448259935
-0.32 2.81000767974183
-0.32 8.41136386459693
-0.32 1.31732914736494
-0.32 3.50523523250595
-0.32 5.13712069401518
-0.32 7.09784516012296
-0.32 3.79199902759865
-0.32 3.45124143613502
-0.32 19.5497866728343
-0.32 12.8058322991245
};
\path [draw=dimgray84, fill=peru21713894]
(axis cs:-0.24,0.00734716746956105)
--(axis cs:-0.08,0.00734716746956105)
--(axis cs:-0.08,0.016003965260461)
--(axis cs:-0.24,0.016003965260461)
--(axis cs:-0.24,0.00734716746956105)
--cycle;
\addplot [dimgray84, forget plot]
table {%
-0.16 0.00734716746956105
-0.16 0.0014710096642374
};
\addplot [dimgray84, forget plot]
table {%
-0.16 0.016003965260461
-0.16 0.028856579773128
};
\addplot [dimgray84, forget plot]
table {%
-0.2 0.0014710096642374
-0.12 0.0014710096642374
};
\addplot [dimgray84, forget plot]
table {%
-0.2 0.028856579773128
-0.12 0.028856579773128
};
\addplot [black, mark=o, mark size=2, mark options={solid,fill opacity=0,draw=dimgray84}, only marks, forget plot]
table {%
-0.16 0.0309758127667009
-0.16 0.039395844284445
-0.16 0.0429825571365654
-0.16 0.0340663902461528
-0.16 0.0290691467933356
-0.16 0.0548090088181197
-0.16 0.0843509363010525
-0.16 0.0575564169324934
-0.16 0.0359818440862
-0.16 0.0351577832363545
-0.16 0.0428436201065778
-0.16 0.0583449120633304
-0.16 0.0478084518574178
-0.16 0.0337549372576177
-0.16 0.0546475179493427
-0.16 0.0342542557977139
-0.16 0.0292806768789887
-0.16 0.0349450702778995
-0.16 0.0606946199201047
-0.16 0.0364966255612671
-0.16 0.0410255820490419
-0.16 0.0389387211762368
-0.16 0.0357239455915987
-0.16 0.0508759362623095
-0.16 0.0295008853077888
-0.16 0.031959512270987
-0.16 0.0312205478549003
-0.16 0.0330214007757604
-0.16 0.0366323491558432
-0.16 0.0383674731478095
};
\path [draw=dimgray84, fill=peru21713894]
(axis cs:0.76,0.113058291841298)
--(axis cs:0.92,0.113058291841298)
--(axis cs:0.92,0.157669330947101)
--(axis cs:0.76,0.157669330947101)
--(axis cs:0.76,0.113058291841298)
--cycle;
\addplot [dimgray84, forget plot]
table {%
0.84 0.113058291841298
0.84 0.0666179896021882
};
\addplot [dimgray84, forget plot]
table {%
0.84 0.157669330947101
0.84 0.196874585313102
};
\addplot [dimgray84, forget plot]
table {%
0.8 0.0666179896021882
0.88 0.0666179896021882
};
\addplot [dimgray84, forget plot]
table {%
0.8 0.196874585313102
0.88 0.196874585313102
};
\addplot [black, mark=o, mark size=2, mark options={solid,fill opacity=0,draw=dimgray84}, only marks, forget plot]
table {%
0.84 0.226019851242503
};
\path [draw=dimgray84, fill=peru21713894]
(axis cs:1.76,5.21867410660101)
--(axis cs:1.92,5.21867410660101)
--(axis cs:1.92,7.46654969795297)
--(axis cs:1.76,7.46654969795297)
--(axis cs:1.76,5.21867410660101)
--cycle;
\addplot [dimgray84, forget plot]
table {%
1.84 5.21867410660101
1.84 2.75931388108681
};
\addplot [dimgray84, forget plot]
table {%
1.84 7.46654969795297
1.84 7.58479055948555
};
\addplot [dimgray84, forget plot]
table {%
1.8 2.75931388108681
1.88 2.75931388108681
};
\addplot [dimgray84, forget plot]
table {%
1.8 7.58479055948555
1.88 7.58479055948555
};
\addplot [black, mark=o, mark size=2, mark options={solid,fill opacity=0,draw=dimgray84}, only marks, forget plot]
table {%
1.84 20.0944223081072
};
\path [draw=dimgray84, fill=darkseagreen117191113]
(axis cs:-0.08,0.00670494828373185)
--(axis cs:0.08,0.00670494828373185)
--(axis cs:0.08,0.0141172985080629)
--(axis cs:-0.08,0.0141172985080629)
--(axis cs:-0.08,0.00670494828373185)
--cycle;
\addplot [dimgray84, forget plot]
table {%
0 0.00670494828373185
0 0.0013204648159444
};
\addplot [dimgray84, forget plot]
table {%
0 0.0141172985080629
0 0.0247371051460504
};
\addplot [dimgray84, forget plot]
table {%
-0.04 0.0013204648159444
0.04 0.0013204648159444
};
\addplot [dimgray84, forget plot]
table {%
-0.04 0.0247371051460504
0.04 0.0247371051460504
};
\addplot [black, mark=o, mark size=2, mark options={solid,fill opacity=0,draw=dimgray84}, only marks, forget plot]
table {%
0 0.0536210935562849
0 0.0455659674480557
0 0.0465836218558251
0 0.0258979379199445
0 0.0275317147374153
0 0.0270288660190999
0 0.111338030081242
0 0.0344643815420568
0 0.0263282165862619
0 0.0257759564556181
0 0.0263590020127594
0 0.029207010474056
0 0.114646213129163
0 0.0266264627687633
0 0.133420317061245
0 0.0257967656478285
0 0.063892110157758
0 0.0277161694131791
0 0.0706993457861244
0 0.0304002902470529
0 0.0341076740063726
0 0.0391403056681156
0 0.129879016056657
0 0.0767357799224555
0 0.0949600432068109
0 0.0467860613018274
0 0.0377010715194046
0 0.0286807938478887
0 0.0428135649301111
0 0.173108574561775
0 0.0291259340010583
0 0.0325142444111406
0 0.0729220287874341
0 0.0406011242419481
0 0.039365014154464
0 0.0685595159418881
0 0.0493802280165255
0 0.0422349452972412
0 0.0273016995750367
0 0.0271273413673043
0 0.0266137507744133
0 0.0470162506215274
0 0.0261966892518103
0 0.086756950803101
};
\path [draw=dimgray84, fill=darkseagreen117191113]
(axis cs:0.92,11.906882261547)
--(axis cs:1.08,11.906882261547)
--(axis cs:1.08,20.026265696312)
--(axis cs:0.92,20.026265696312)
--(axis cs:0.92,11.906882261547)
--cycle;
\addplot [dimgray84, forget plot]
table {%
1 11.906882261547
1 3.95808663095037
};
\addplot [dimgray84, forget plot]
table {%
1 20.026265696312
1 28.4920856157939
};
\addplot [dimgray84, forget plot]
table {%
0.96 3.95808663095037
1.04 3.95808663095037
};
\addplot [dimgray84, forget plot]
table {%
0.96 28.4920856157939
1.04 28.4920856157939
};
\path [draw=dimgray84, fill=indianred199109109]
(axis cs:0.0799999999999999,0.106594913018246)
--(axis cs:0.24,0.106594913018246)
--(axis cs:0.24,0.117922946810722)
--(axis cs:0.0799999999999999,0.117922946810722)
--(axis cs:0.0799999999999999,0.106594913018246)
--cycle;
\addplot [dimgray84, forget plot]
table {%
0.16 0.106594913018246
0.16 0.0905358716845512
};
\addplot [dimgray84, forget plot]
table {%
0.16 0.117922946810722
0.16 0.134411989400784
};
\addplot [dimgray84, forget plot]
table {%
0.12 0.0905358716845512
0.2 0.0905358716845512
};
\addplot [dimgray84, forget plot]
table {%
0.12 0.134411989400784
0.2 0.134411989400784
};
\addplot [black, mark=o, mark size=2, mark options={solid,fill opacity=0,draw=dimgray84}, only marks, forget plot]
table {%
0.16 0.0858727321028709
0.16 0.0855136004587014
0.16 0.0875271006176869
0.16 0.0875803263237079
0.16 0.0865102844933668
0.16 0.0857456531375646
0.16 0.0821034604062636
0.16 0.0883783549070358
0.16 0.0810442728300889
0.16 0.0866780206561088
0.16 0.0867269473771254
0.16 0.0891289630283912
0.16 0.0879806050409873
0.16 0.0867695988466342
0.16 0.0872577652335167
0.16 0.0889238597204287
0.16 0.0876092128455638
0.16 0.0893326171984275
0.16 0.0799844898283481
0.16 0.0787164705495039
0.16 0.0875431771079699
0.16 0.0882449795802434
0.16 0.0848105028271675
0.16 0.167169097810984
0.16 0.296551653494438
0.16 0.185926345487436
0.16 0.14308217416207
0.16 0.181506453081965
0.16 0.135954804718494
0.16 0.155331154664358
0.16 0.138198221723239
0.16 0.157449054842194
0.16 0.147079659625888
0.16 0.181544989347458
0.16 0.165998588626583
0.16 0.157665833830833
0.16 0.211310898885131
0.16 0.136157508318623
0.16 0.186892301465074
0.16 0.137743590399623
0.16 0.145681707809369
0.16 0.171230843290687
0.16 0.144670408839981
0.16 0.14557481619219
0.16 0.227404676998655
0.16 0.183375282213092
0.16 0.135107697298129
0.16 0.153337599088748
0.16 0.142070634911458
};
\path [draw=dimgray84, fill=indianred199109109]
(axis cs:1.08,1.77062407632669)
--(axis cs:1.24,1.77062407632669)
--(axis cs:1.24,2.83666957045595)
--(axis cs:1.08,2.83666957045595)
--(axis cs:1.08,1.77062407632669)
--cycle;
\addplot [dimgray84, forget plot]
table {%
1.16 1.77062407632669
1.16 0.911291380723318
};
\addplot [dimgray84, forget plot]
table {%
1.16 2.83666957045595
1.16 3.71264587280651
};
\addplot [dimgray84, forget plot]
table {%
1.12 0.911291380723318
1.2 0.911291380723318
};
\addplot [dimgray84, forget plot]
table {%
1.12 3.71264587280651
1.2 3.71264587280651
};
\path [draw=dimgray84, fill=indianred199109109]
(axis cs:2.08,291.875387804893)
--(axis cs:2.24,291.875387804893)
--(axis cs:2.24,338.623210176826)
--(axis cs:2.08,338.623210176826)
--(axis cs:2.08,291.875387804893)
--cycle;
\addplot [dimgray84, forget plot]
table {%
2.16 291.875387804893
2.16 290.267973103871
};
\addplot [dimgray84, forget plot]
table {%
2.16 338.623210176826
2.16 348.338838763535
};
\addplot [dimgray84, forget plot]
table {%
2.12 290.267973103871
2.2 290.267973103871
};
\addplot [dimgray84, forget plot]
table {%
2.12 348.338838763535
2.2 348.338838763535
};
\addplot [black, mark=o, mark size=2, mark options={solid,fill opacity=0,draw=dimgray84}, only marks, forget plot]
table {%
2.16 144.184378642589
2.16 461.006676754604
};
\path [draw=dimgray84, fill=lightslategray147116171]
(axis cs:0.24,0.0061810553539544)
--(axis cs:0.4,0.0061810553539544)
--(axis cs:0.4,0.00975948181003325)
--(axis cs:0.24,0.00975948181003325)
--(axis cs:0.24,0.0061810553539544)
--cycle;
\addplot [dimgray84, forget plot]
table {%
0.32 0.0061810553539544
0.32 0.001457498781383
};
\addplot [dimgray84, forget plot]
table {%
0.32 0.00975948181003325
0.32 0.0149294063448905
};
\addplot [dimgray84, forget plot]
table {%
0.28 0.001457498781383
0.36 0.001457498781383
};
\addplot [dimgray84, forget plot]
table {%
0.28 0.0149294063448905
0.36 0.0149294063448905
};
\addplot [black, mark=o, mark size=2, mark options={solid,fill opacity=0,draw=dimgray84}, only marks, forget plot]
table {%
0.32 0.015269245672971
0.32 18.1096389207989
0.32 0.0186256303451955
0.32 0.0197214713320136
0.32 0.0160867095924913
0.32 0.0192236443981528
0.32 0.0155682682059705
0.32 0.0180801945738494
0.32 0.0155066839419305
0.32 0.0155029710382223
0.32 0.0211028065532445
0.32 0.0180883092805743
0.32 0.0151781074702739
0.32 0.0154028967954218
0.32 0.016799986641854
0.32 0.0167776033282279
};
\path [draw=dimgray84, fill=lightslategray147116171]
(axis cs:1.24,0.0476908226652691)
--(axis cs:1.4,0.0476908226652691)
--(axis cs:1.4,0.0716967967649301)
--(axis cs:1.24,0.0716967967649301)
--(axis cs:1.24,0.0476908226652691)
--cycle;
\addplot [dimgray84, forget plot]
table {%
1.32 0.0476908226652691
1.32 0.0243114571397503
};
\addplot [dimgray84, forget plot]
table {%
1.32 0.0716967967649301
1.32 0.0964580938840905
};
\addplot [dimgray84, forget plot]
table {%
1.28 0.0243114571397503
1.36 0.0243114571397503
};
\addplot [dimgray84, forget plot]
table {%
1.28 0.0964580938840905
1.36 0.0964580938840905
};
\path [draw=dimgray84, fill=lightslategray147116171]
(axis cs:2.24,1.62898788647726)
--(axis cs:2.4,1.62898788647726)
--(axis cs:2.4,2.39610649823832)
--(axis cs:2.24,2.39610649823832)
--(axis cs:2.24,1.62898788647726)
--cycle;
\addplot [dimgray84, forget plot]
table {%
2.32 1.62898788647726
2.32 0.93436328911533
};
\addplot [dimgray84, forget plot]
table {%
2.32 2.39610649823832
2.32 2.40685283765197
};
\addplot [dimgray84, forget plot]
table {%
2.28 0.93436328911533
2.36 0.93436328911533
};
\addplot [dimgray84, forget plot]
table {%
2.28 2.40685283765197
2.36 2.40685283765197
};
\addplot [black, mark=o, mark size=2, mark options={solid,fill opacity=0,draw=dimgray84}, only marks, forget plot]
table {%
2.32 7.24691668370118
};
\addplot [dimgray84, forget plot]
table {%
-0.4 0.133921613544226
-0.24 0.133921613544226
};
\addplot [dimgray84, forget plot]
table {%
-0.24 0.010836489778012
-0.08 0.010836489778012
};
\addplot [dimgray84, forget plot]
table {%
0.76 0.137572378230592
0.92 0.137572378230592
};
\addplot [dimgray84, forget plot]
table {%
1.76 6.19084846787155
1.92 6.19084846787155
};
\addplot [dimgray84, forget plot]
table {%
-0.08 0.0099125517532229
0.08 0.0099125517532229
};
\addplot [dimgray84, forget plot]
table {%
0.92 16.7897997110461
1.08 16.7897997110461
};
\addplot [dimgray84, forget plot]
table {%
0.0799999999999999 0.111437018960714
0.24 0.111437018960714
};
\addplot [dimgray84, forget plot]
table {%
1.08 2.38464089979728
1.24 2.38464089979728
};
\addplot [dimgray84, forget plot]
table {%
2.08 303.086978162328
2.24 303.086978162328
};
\addplot [dimgray84, forget plot]
table {%
0.24 0.0081863400526344
0.4 0.0081863400526344
};
\addplot [dimgray84, forget plot]
table {%
1.24 0.0582122245493034
1.4 0.0582122245493034
};
\addplot [dimgray84, forget plot]
table {%
2.24 2.0821574012128
2.4 2.0821574012128
};

\addplot [mark=triangle*,mark options={scale=2, fill=steelblue89124191}] coordinates {
    (0.68, 1000)
    (1.68, 1000)
}; %
\addplot [mark=triangle*,mark options={scale=2, fill=darkseagreen117191113}] coordinates {
    (2, 1000)
}; %
\end{axis}

\end{tikzpicture}